\documentclass[12pt]{article}
\usepackage[utf8]{inputenc}

\usepackage[top=1in, bottom=1in, left=1in, right=1in]{geometry}
\usepackage{graphicx} 
\usepackage{subfig}
\usepackage{amsthm}
\usepackage[authoryear]{natbib}
\usepackage{ulem}
\usepackage{color}
\usepackage{lineno}  
\usepackage{hyperref}
\usepackage{amsmath}
\usepackage{amssymb}
\usepackage{indentfirst} 

\numberwithin{equation}{section}
\newcommand\keywords[1]{\textbf{Keywords}: #1}

\title{Stability analysis of spatial perturbed elliptic restricted \\ 3-body problem with double-averaging }
\author{\small{Yan Luo$^2$, Kaicheng Sheng$^1$\footnote{Corresponding author. E-mail addresses:  {\it k.sheng@sdu.edu.cn} (K.Sheng)}} \\
\small{$^1$ School of Mathematics, Shandong University, Jinan, China }\\
\small{$^2$ Research Centre for Mathematics and Interdisciplinary Sciences,}\\ \small{Shandong University, Qingdao, China }
}

\date{}

\begin{document}

%\linenumbers

\maketitle

\begin{abstract}
This paper investigates the secular motion of a massless asteroid within the framework of the double-averaged elliptic restricted three-body problem. By employing Poincar\'e variables, the stability properties of asteroid orbits in the presence of spatial perturbations were analysed. The study reveals that periodic orbits identified in the planar configuration maintain stability in the spatial perturbed problem across a wide range of parameter values. These findings, supported by numerical simulations, contribute to a deeper understanding of asteroid dynamics and have implications for studying exoplanetary systems with highly eccentric host stars. 
\end{abstract}




\section{Introduction}

The study of the stability of celestial bodies within gravitational systems is a foundational aspect of celestial mechanics. The study of the 3-body problem formed the foundation for understanding the complex dynamic behaviours observed in natural and artificial systems. The restricted 3-body problem plays an important model role in the 3-body problem model. This paper focuses on the double-averaged spatial perturbed elliptic restricted 3-body problem (ER3BP) involving a star, a planet, and an asteroid. The study is under the assumption that the mass of the planet is much smaller than the mass of the star. The averaging over fast phases corresponds to the motions of the star-planet system and the asteroid. The classical restricted 3-body problem has been instrumental in developing our understanding of orbital mechanics, particularly in cases where the orbits of the primary bodies are assumed to be circular. However, in reality, most celestial bodies, including stars, planets and asteroids, follow elliptical orbits, which introduces significant complexities to the problem. These complexities necessitate more advanced mathematical techniques and have led to the development of the ER3BP, where the primary bodies follow elliptical orbits. 

Double-averaging method simplifies the complex gravitational interactions, resulting in a simplified system that retains the essential dynamical features of the original problem while being more tractable for analysis over long timescales. The seminal work of \citet{akse} on the double-averaged elliptical restricted 3-body problem laid the foundation for understanding the long-term evolution of orbits in such systems. His work demonstrated how the double-averaging method could be used to reduce the complexity of the problem, making it possible to derive analytical expressions for the evolution of orbital elements over time. 

The significance of stability in celestial mechanics, particularly in the context of Hamiltonian systems, has been explored by \citet{Arnold1961}, who established fundamental results on the stability of equilibrium positions. This work is further elaborated in the comprehensive treatment of classical mechanics by \citet{akn}. Moreover, the secular evolution of orbits, especially in hierarchical systems, has been investigated by \citet{lidov} and \citet{kozai}, whose work on the Lidov-Kozai mechanism has profoundly influenced our understanding of orbital dynamics in N-body systems. The relevance of the Lidov-Kozai mechanism has been further underscored by recent studies, such as those by \citet{Katz2011} and \citet{Lithwick2011}, who explored its effects in systems with eccentric perturbers. 

The stability of planar orbits in double-averaged circular problems was established by \citet{neish}.  His study indicates that planar orbits also remain stable in the linear approximation of the double-averaged elliptic problem with a sufficiently small eccentricity of the perturber's orbit. The stability of periodic solutions in the restricted 3-body problem, particularly in the presence of elliptical orbits, has also been explored by \citet{Leontovich} and \citet{Moser1968}, whose work on Hamiltonian systems with two degrees of freedom provided crucial insights into the resonance phenomena that often govern the stability of such systems. \citet{Harrington1968} and \citet{zig} analyzed the dynamical evolution of the 3-body problem, while more recent studies, such as those by \citet{naoz2013} and \citet{Sidorenko2018}, have extended these ideas to modern exoplanetary systems. The double-averaged ER3BP specifically benefits from these approaches, offering insights into the stability conditions that govern the long-term behaviour of small bodies in elliptical orbits.

In recent years, there has been renewed interest in the study of the double-averaged ER3BP, particularly in the context of exoplanetary systems. The discovery of numerous exoplanets with highly eccentric orbits has prompted researchers to revisit the problem of stability in such systems. Studies by \citet{HL} and \citet{Lei} have focused on the linear stability of the inner case of the double-averaged spatial elliptic restricted 3-body problem, providing new insights into the conditions under which stability can be achieved in these systems. A completed numerical analysis, including all possible situations for the evolution of planar orbits in the double-averaged ER3BP, can be found in the work by \citet{vash_planar}.

The investigation of equilibria within the double-averaged ER3BP, as studied by \citet{Palacian2006}, and the analysis of apsidal alignment by \citet{NSS} analyzed both the linear and nonlinear stability of apsidal alignment in the spatial double-averaged ER3BP, further illustrated the rich dynamical behaviour that emerges in such systems. Linear stability of the equilibria and periodic orbits of the asteroid in spatial double-averaged ER3BP in the inner case with arbitrary inclination is studied in \citet{HLSY}. 

This paper aims to provide a comprehensive analysis of the stability of the spatial perturbed double-averaged ER3BP, building on the foundational work of \citet{akse}, \citet{vashkovyak, vash_planar}, \citet{NSS} and others. By combining analytical methods with numerical simulations, the long-term stability of the asteroid's orbits in double-averaged ER3BP is investigated in linear approximation. The results have important implications for both theoretical and practical aspects of celestial mechanics, providing new insights into the stability of orbits in a wide range of astrophysical systems. Through this work, we hope to contribute to the ongoing exploration of the dynamic and ever-evolving nature of the cosmos.

A notable characteristic of many exoplanets is their orbits, which often exhibit large eccentricities and inclinations, deviating significantly from the near-circular and coplanar trajectories observed in the solar system \citep{skdc,subaru}. Our research holds the potential to make substantial contributions to the domain of exoplanetary dynamics \citep{shev}. Furthermore, the secular evolution of motions within the non-restricted three-body problem has been extensively studied through the application of averaging techniques \citep{Harrington1968,zig,Mi2004}, one can extend our work into the non-restricted 3-body problem.

\section{Statement of the problem}

\begin{figure}[htbp] 
\centering
\includegraphics[width=0.5\textwidth]{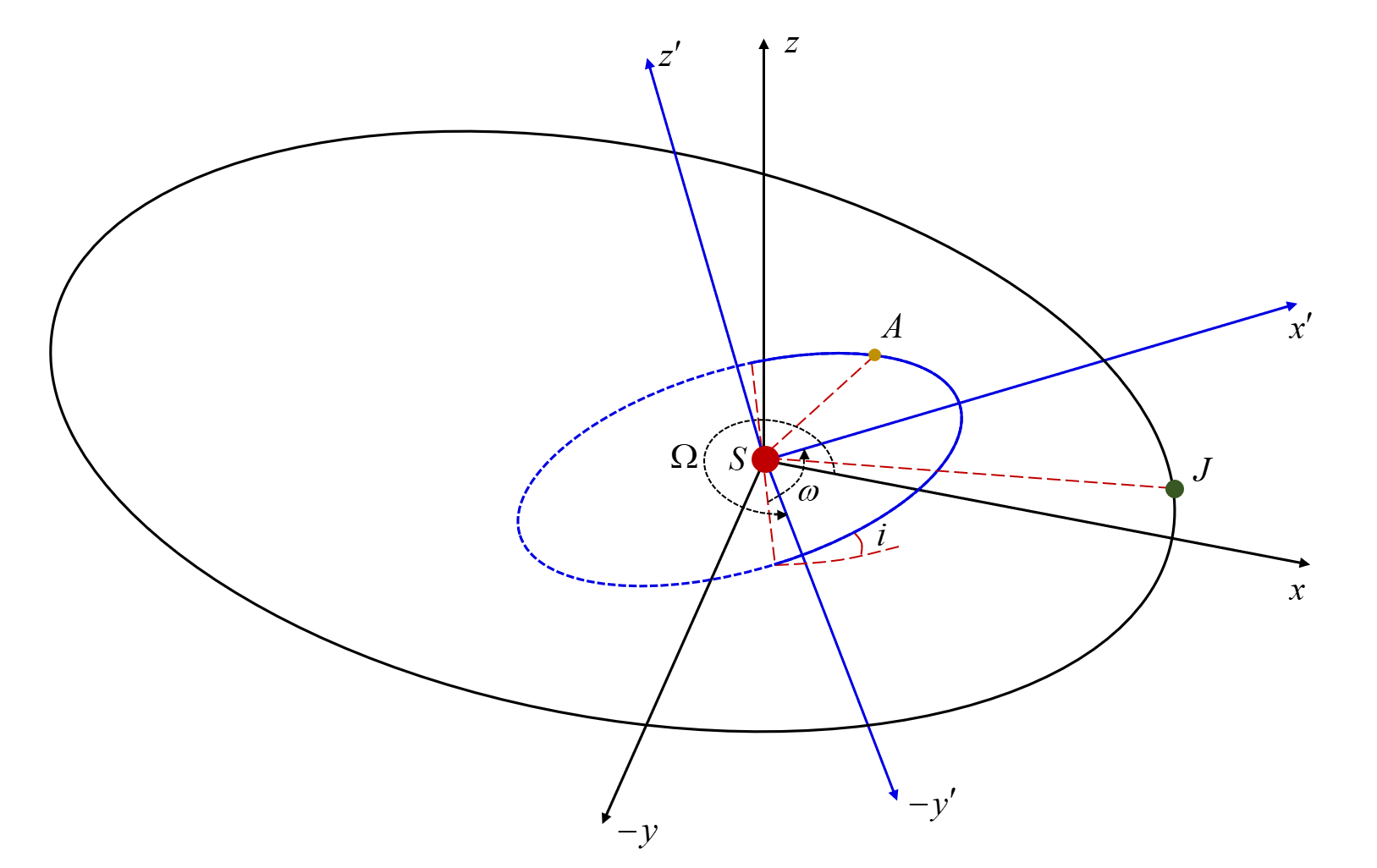} 
\caption{Coordinate frames.}
\label{fig:tu77} 
\end{figure} 

The considered spatial perturbed elliptic restricted three-body problem involves a star $S$, a planet $J$, and an asteroid $A$ \citep{BC}. Adopting a coordinate system analogous to that in \citet{NSS}, the star is positioned at the origin $O$ of a Cartesian frame $Oxyz$. The $ Oxy$ plane of this system is defined by the orbital plane of the star-planet system. Consequently, the coordinates of the planet and the asteroid are $(x_{J},y_{J},0)$ and $(x,y,z)$, respectively. A rotating Cartesian frame $Ox'y'z'$ is introduced, where the $ Ox'y'$ plane aligns with the asteroid's osculating orbital plane. In this frame, the asteroid's coordinates are $(x', y',0)$. The asteroid’s orbital motion is characterized by standard osculating elements: semi-major axis ($a$), mean anomaly ($l$), eccentricity ($e$), argument of periapsis ($\omega$), inclination ($i$), and longitude of the ascending node ($\Omega$). Refer to Fig. \ref{fig:tu77} for a visual representation of the system geometry. According to the rotation of the Cartesian frame, the transformation between frames $Oxyz$ and $Ox'y'z'$ is 
\begin{equation}
\left[\begin{array}{ccc}
x & y & z
\end{array}\right]^T=\mathcal{T}_{\Omega}\times\mathcal{T}_{i}\times\mathcal{T}_{\omega}\times\left[\begin{array}{ccc}
x' & y' & 0
\end{array}\right]^T,
\end{equation}
where 
\begin{equation}
\begin{aligned}
\mathcal{T}_{\Omega}&=\left[\begin{array}{ccc}
\cos\left(\Omega\right) & -\sin\left(\Omega\right) & 0 
\\
 \sin\left(\Omega\right) & \cos\left(\Omega\right) & 0 
\\
 0 & 0 & 1 
\end{array}\right],\, 
\\
\mathcal{T}_{i}&=\left[\begin{array}{ccc}
1 & 0 & 0 
\\
 0 & \cos\! \left(i\right) & -\sin\! \left(i\right) 
\\
 0 & \sin\! \left(i\right) & \cos\! \left(i\right) 
\end{array}\right], \, 
\\
\mathcal{T}_{\omega}&=\left[\begin{array}{ccc}
\cos\left(\omega\right) & -\sin\left(\omega\right) & 0 
\\
 \sin\left(\omega \right) & \cos\left(\omega\right) & 0 
\\
 0 & 0 & 1 
\end{array}\right].
\end{aligned}
\end{equation}
Consequently, the expressions of $x$, $y$ and $z$ are
\begin{equation}
\label{coord_transform}
\begin{aligned}
x &=\left( \cos\Omega
\cos\omega -\cos  i\sin\Omega
  \sin\omega  \right) x'+\left( -\cos\Omega\sin  
\omega  -\cos i\sin\Omega \cos
\omega  \right) y',
\\ 
 y &=\left( \sin 
\Omega  \cos\omega +\cos i\cos\Omega
 \sin\omega  \right) x' +\left( -\sin\Omega\sin
\omega +\cos  i\cos\Omega
  \cos\omega  \right) y' ,
\\ 
 z &=\left(\sin  i
  \sin\omega\right) x'+ \left(\sin i \cos\omega\right) y'. 
\end{aligned}
\end{equation}

It follows from \citet{shev}, the planet moves in a prescribed elliptic orbit:
\begin{equation}
\begin{aligned}
\label{coord_J}
 {{x}_{J}}&={{a}_{J}}\left( \cos{{E}_{J}} -{{e}_{J}} \right),
 \\
  {{y}_{J}}&={{a}_{J}}\sqrt{1-{{e}_{J}}^{2}}\, \sin  {{E}_{J}} ,
  \\
  l_J&=E_J-e_J\sin E_J. 
\end{aligned}
\end{equation}
Here $a_J$, $e_J$, $E_J$, $l_J$ are the semi-major axis, the eccentricity,  the eccentric anomaly, and the mean anomaly of the planet's orbit.  We put $a_J=1$ for convenience in the following.

\section{Hamiltonian of the system}

We introduce the canonical Delaunay elements $(l, g, h, L, G, H)$, where $g\left(\equiv \omega\right)$ and $h\left(\equiv \Omega\right)$ are the argument of pericenter and the ascending node of the asteroid. The elements $L=\sqrt{(1-\mu)a}$, $G=L\sqrt{1-e^{2}}$, and $H=G\cos i$ correspond to the Keplerian energy, total angular momentum, and $z$-component of angular momentum, respectively \citep{BC}. To facilitate the dynamical analysis of the asteroid, canonical Poincar\'e variables are denoted by 

\begin{equation}
\label{P_elements}
\begin{aligned}
 &{{p}_{1}}=L, &&{{q}_{1}}=l+g+h,
  \\ 
  &{{p}_{2}}=\sqrt{2\left( L-G \right)}\cos\left( g+h \right),
  &&{{q}_{2}}=-\sqrt{2\left( L-G \right)}\sin \left( g+h \right),
  \\ 
 &{{p}_{3}}=\sqrt{2\left( G-H \right)}\cos  h , 
  &&{{q}_{3}}=-\sqrt{2\left( G-H \right)}\sin  h.
\end{aligned} 
\end{equation}

Assuming the planet and star have masses $\mu$ and $1-\mu$, respectively, such that the total system mass equals unity. $\mu \ll 1$. The Hamiltonian of the asteroid is given by:
\begin{equation}
\label{F}
F=-{\frac {\left( 1-\mu \right) ^{2}}{2{L}^{2}}}+\mu U -\mu (x\ddot x_J+y\ddot y_J), 
\end{equation}
where
\begin{equation}
\label{U}
U=-V=-{\frac {1}{\sqrt{(x-x_{J})^{2}+(y-y_{J})^{2}+{z}^{2}}}}
\end{equation}
is the perturbing gravitational potential which is expressed in terms of the asteroid's coordinates $(x,y,z)$. These coordinates can be transformed into Poincar\'e variables with equations (\ref{coord_transform}) and (\ref{P_elements}). The asteroid's motion in its elliptic orbit is governed by the equations:
\begin{equation}
\begin{aligned}
%\label{coord_A}
 {{x'}}&={{a}}(\cos E-e),
 \\
 {{y'}}&={{a}}\sqrt{1-{{e}}^{2}} \sin  {{E}} ,
 \\
 l&=E-e\sin E,
 \end{aligned} 
\end{equation}
where $E$ denotes the eccentric anomaly \citep{shev}. The planet's coordinates, $x_J$ and $y_J$, are considered prescribed functions of time.

The double-averaged Hamiltonian $\bar F$ of the asteroid is defined as:
\begin{equation}
%\label{bar_F}
\bar F =\frac{1}{(2\pi)^2}\iint_{[0,2\pi]^2} F dldl_J = -{\frac {\left( 1-\mu \right) ^{2}}{2{L}^{2}}} + \mu {\bar U}= -{\frac {\left( 1-\mu \right) ^{2}}{2{L}^{2}}} - \mu {\bar V},     
\end{equation}
where 
\begin{equation}
\label{bar_U}
\bar U =\frac{1}{(2\pi)^2}\iint_{[0,2\pi]^2} U dldl_J,
\end{equation}
and $\mu \bar V=-\mu \bar U$  is the double-averaged force function of gravity of the planet. 

Performing the double-averaging over the mean anomalies of the planet and the asteroid, the Hamiltonian becomes independent of $q_1$. Consequently, the conjugate momentum $p_1$ (equivalent to $L$) constitutes a first integral of the system. The first term in the double-averaged Hamiltonian, $\bar F$, thus remains constant. The dynamics of the remaining variables, $\left(p_2, q_2\right)$, $\left(p_3, q_3\right)$, are therefore governed by a two-degree-of-freedom (2-DOF) Hamiltonian system with Hamiltonian $\mu \bar U$.

By introducing the slow time variable $\tau = \mu t$, the Hamiltonian transforms to $\bar{U}$. Notably, $\bar{U}$ depends on the planet's eccentricity, $e_J$, as parameters, and the ratio between the semi-major axis $a$ of the asteroid and $a_J=1$ of the planet. 

Taking $i=0$, the spatial perturbed ER3BP reduces to a planar problem on the invariant plane $p_3=0$, $q_3=0$. This planar system is described by a one-degree-of-freedom (1-DOF) Hamiltonian, $\bar{R}_{\Theta} (p_2,q_2)$, where $\bar{R}_{\Theta} (p_2,q_2)=\bar U(p_2,q_2,p_3 =0,q_3 =0)$ is independent of $p_3$ and $q_3$. Following \citet{akse} and \citet{NSS}, the Hamiltonian of the double-averaged planar ER3BP is given by:
\begin{equation}
%\label{bar_R}
\bar{R}_{\Theta} =\frac{1}{(2\pi)^2}\iint_{[0,2\pi]^2}Rdldl_J,
\end{equation}
where
\begin{equation}
%\label{bar_R}
{R}_{\Theta} =-{\frac {1}{\sqrt{(x-x_{J})^{2}+(y-y_{J})^{2}}}}.
\end{equation}

\begin{figure}[htbp]
  \centering
{\includegraphics[width=0.5\textwidth]{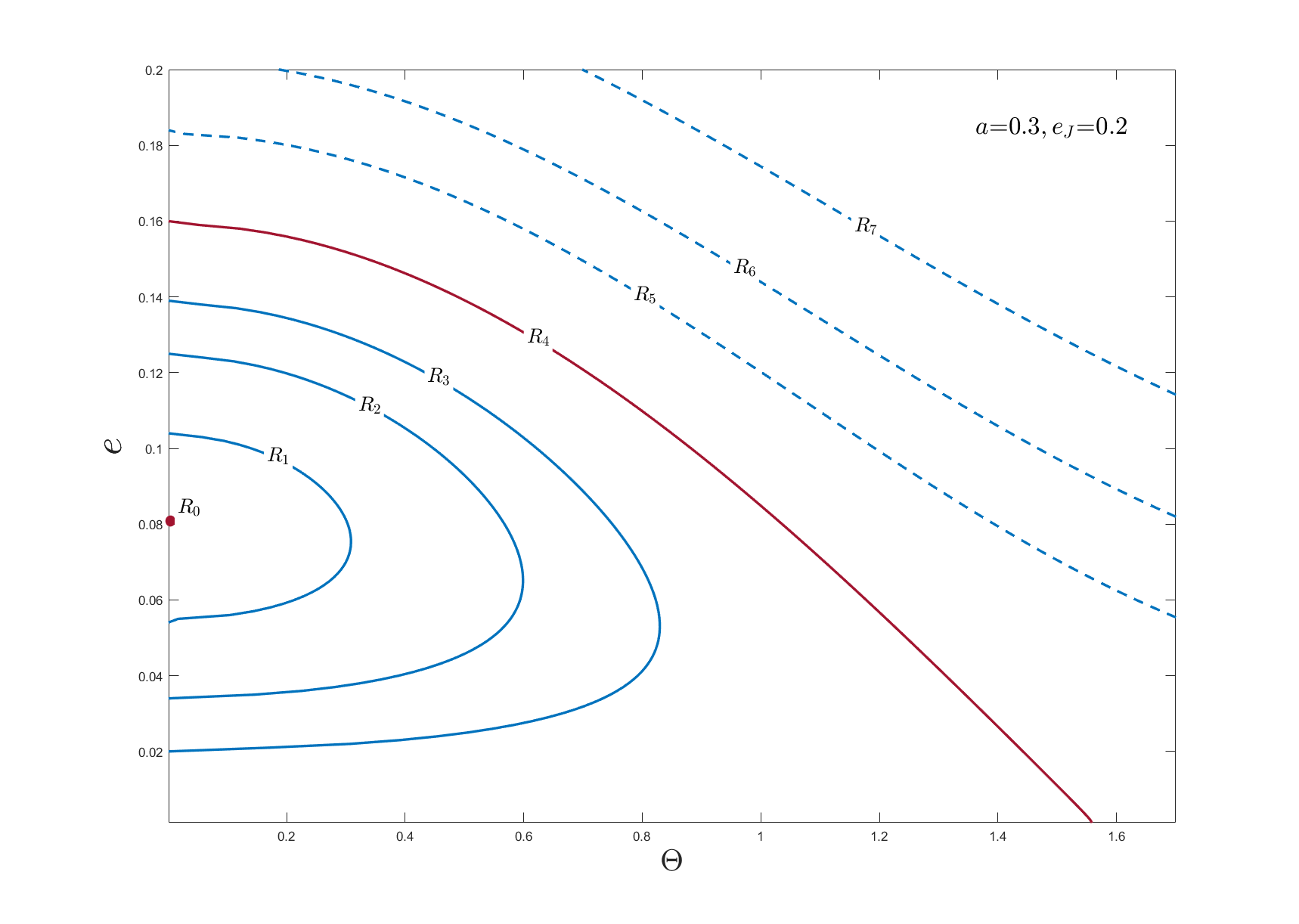}}
\caption{Trajectories of $\Theta$, $e$ in double-averaged planar ER3BP.}
\label{fig:Rphase123} 
\end{figure}

Numerical investigations by \citet{vash_planar} have demonstrated the existence of stationary solutions (equilibria) 
\begin{equation}
\label{ss}
q_2=0,\quad p_2=p_{2*},
\end{equation}
for some domains in the plane of parameters $a$, $e_J$ in the double-averaged planar ER3BP. Periodic orbits, separatrix and orbits of circulating are included as well. Fixing the parameter $a$ and $e_J$ in $\bar{R}_{\Theta}$, one can obtain the figure of $\Theta$ and $e$. See Fig. \ref{fig:Rphase123}. 

$p_{2*}$ is the root of the equation
\begin{equation}
\label{for_e_*}
\left.\frac{\partial \bar R}{\partial p_2}\right|_{q_2=0}=0.    
\end{equation}
The value of the eccentricity at the equilibrium~(\ref{ss}) is provided by the equation
\begin{equation}
e_* = \sqrt{1-\left(1 - \frac{p^2_{2*}}{2\sqrt{a}}\right)^2}.
\label{for_e_*_1}
\end{equation}

\begin{figure}[htbp]
  \centering
\subfloat[]{\includegraphics[width=0.45\textwidth]{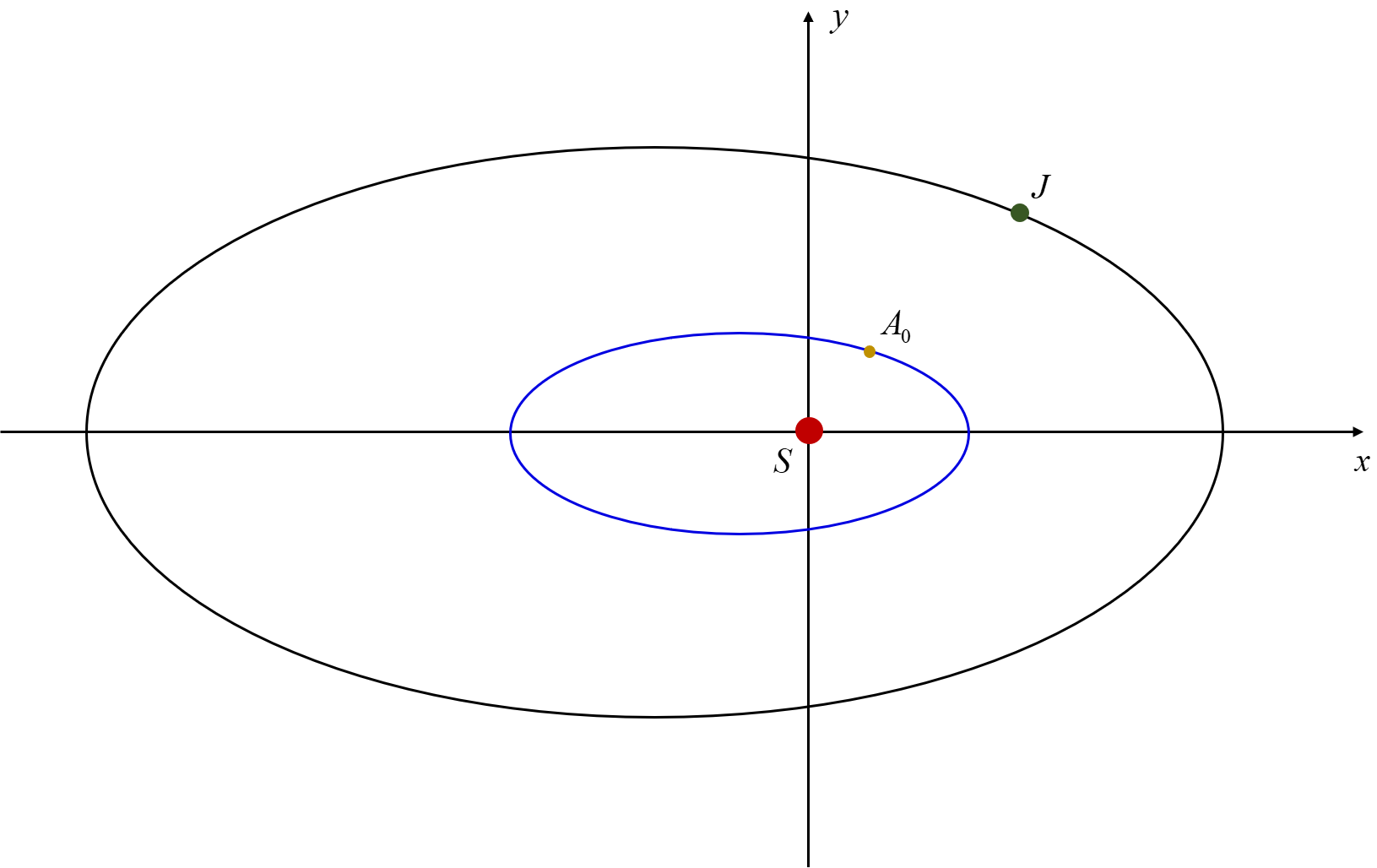}\label{planer1}}  
\subfloat[]{\includegraphics[width=0.45\textwidth]{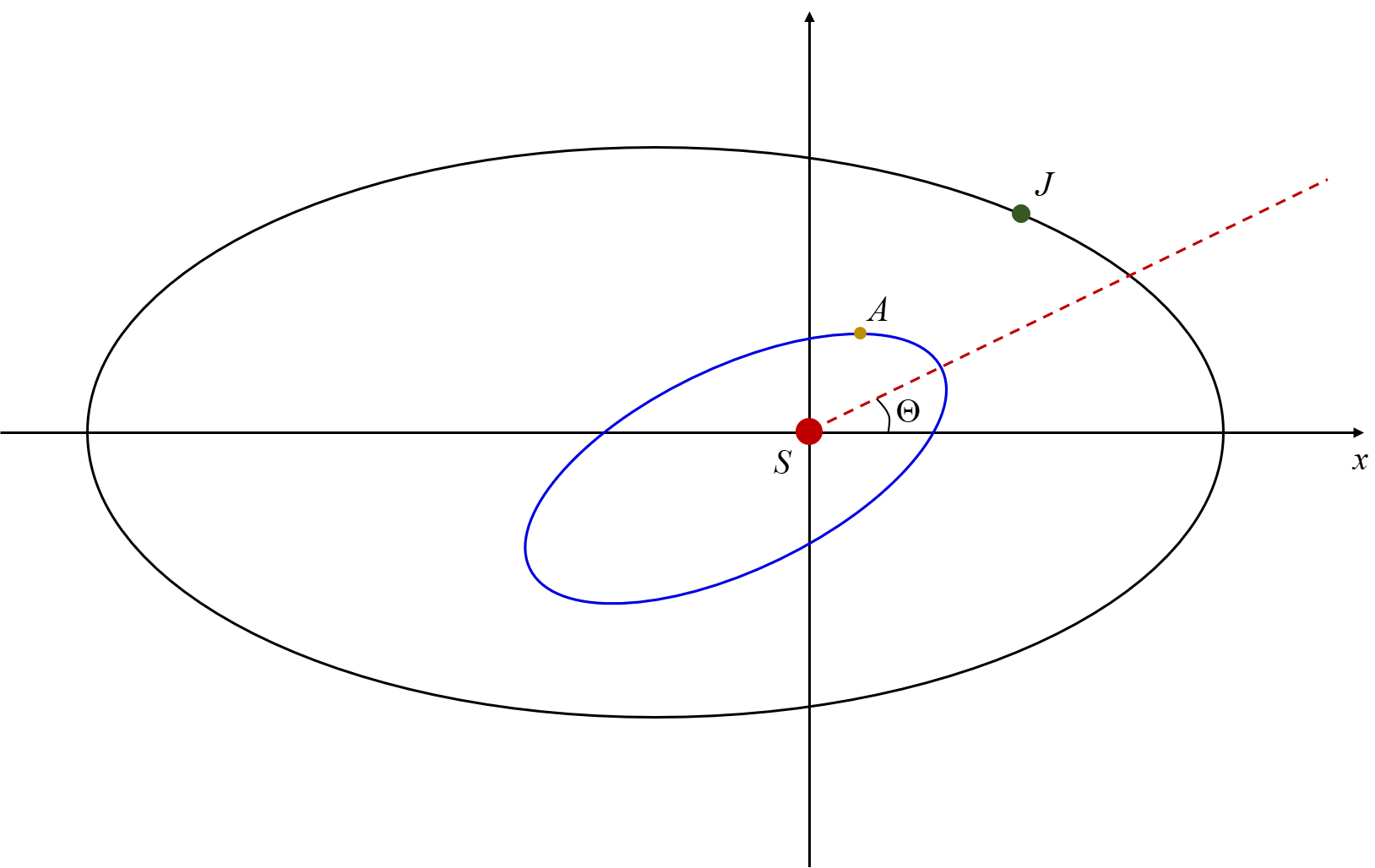}\label{planer2}}
\caption{Case of equilibria (apsidal alignment) and periodic orbits.}
%\label{fig:alignment} 
\end{figure}

As detailed in \citet{vash_planar}, the equilibria defined by equation (\ref{ss}) correspond to an apsidal alignment scenario where $\omega+\Omega=0$. (Fig. \ref{planer1}) The study also identifies periodic orbits in terms of $\Theta=\omega+\Omega$ and the asteroid's eccentricity ($e$) surrounding these equilibria. (Fig. \ref{planer2}) %According to \citet{akse}, the approximate Hamiltonian of the double-averaged planar ER3BP is given by
%\begin{equation}
%\bar{R}_{\Theta} =\frac{1}{a} \left[\frac{1}{8} a^{3} \left(2+3 e^{2}\right) \left(1-e_{J}^{2}\right)^{-\frac{3}{2}}-\frac{15}{64} a^{4} \left(4 e+3 e^{3}\right) e_{J}\left(1-e_{J}^{2}\right)^{-\frac{5}{2}} \cos\left(\Theta\right)\right]
%\end{equation}

To comprehensively assess the linear stability of asteroid orbits in the spatial perturbed ER3BP, it is essential to consider non-equilibrium cases (i.e.,  $\Theta \neq 0$), especially to consider the cases of periodic orbits. The small perturbation case is considered in this paper, which means the inclination $i$ is regarded as a small perturbation. In this case, the formula of $x$, $y$ and $z$ can be obtained with $i=0$ and $\omega+\Omega=\Theta$:
\begin{equation}
\label{coord_transform2'}
\begin{aligned}
x &=\cos(\Theta) x'-\sin(\Theta) y',
\\ 
y &=\sin(\Theta) x'+\cos(\Theta) y',
\\
z &=0.  
 \end{aligned}
\end{equation}

For small $i$, proper approximations of formulas (\ref{P_elements}) are obtained:
\begin{equation}
\label{expand_i}
\begin{aligned}
&\cos \left(\Omega \right) \approx \frac{p_{3}}{\sqrt{G \cdot i^2/2}}, && \sin \left(\Omega \right) \approx -\frac{q_{3}}{\sqrt{G \cdot i^2/2}}, 
\\
&\cos\left(\omega \right)\approx\frac{p_{2} p_{3}-q_{2} q_{3}}{\sqrt{G\cdot i^{2}}\, \sqrt{p_{2}^{2}+q_{2}^{2}}}, && \sin\left(\omega \right)\approx-\frac{p_{2} q_{3}+p_{3} q_{2}}{\sqrt{G\cdot i^{2}}\, \sqrt{p_{2}^{2}+q_{2}^{2}}}.
 \end{aligned}
\end{equation}
Substituting (\ref{expand_i}) into (\ref{coord_transform}) and then taking in (\ref{U}), the function $U=U_{\Theta}$ with increasing order of $p_{3}$ and $q_{3}$ can be written as
\begin{equation}
\label{quadratic2}
U_{\Theta}=R_{\Theta}+ W_{\Theta} +O(p_3^4+ q_3^4), \quad W_{\Theta}= \frac{1}{2}\left(A_{\Theta} p_3^2 +2B_{\Theta, } p_3 q_3+C_{\Theta} q_3^2 \right) 
%{= \frac{1}{2} \begin{pmatrix} p_3 & q_3 \end{pmatrix}
%\begin{pmatrix} A_{\Theta} & B_{\Theta} \\ B_{\Theta} & C_{\Theta} \end{pmatrix}
%\begin{pmatrix} p_3 \\ q_3 \end{pmatrix}
\end{equation}
where 
\begin{equation}
\label{ABCR0}   
\begin{aligned}
A_{\Theta}&=\,\frac{y{{y}_{J}}}{G {\left[ {\left( x-{{x}_{J}} \right)}^{2}+{{\left( y-{{y}_{J}} \right)}^{2}} \right]}^{{3}/{2}}}, 
 \\ 
B_{\Theta}&=\,\frac{\left( x{{y}_{J}}+y{{x}_{J}} \right)}{2G {{\left[ {{\left( x-{{x}_{J}} \right)}^{2}}+{{\left( y-{{y}_{J}} \right)}^{2}} \right]}^{{3}/{2}}}}, 
\\
C_{\Theta}&=\, \frac{x{{x}_{J}}}{ G {\left[ {{\left( x-{{x}_{J}} \right)}^{2}}+{{\left( y-{{y}_{J}} \right)}^{2}} \right]^{{3}/{2}}}}.
\end{aligned}
\end{equation}
Here all quantities are calculated at $i=0$ and $\Theta=\omega+\Omega$. Calculations are similar to the procedure in \citet{NSS}.

%We consider the double average value of the coefficients in the quadratic part of  $p_{3}$ and $q_{3}$. 
Averaging $W_{\Theta}$ over the mean anomaly of the asteroid $l$ and the mean anomaly of the planet $l_J$, the averaged value of $W_{\Theta}$ is
\begin{equation}
\label{{W}_{in}_0}
\bar{W}_{\Theta}=\frac{1}{2}\left(\bar{A}_{\Theta}\,{{p}_{3}}^{2}+2\bar{B}_{\Theta} p_3 q_3+\bar{C}_{\Theta}\, {{q}_{3}}^{2}\right),
\end{equation}
where 
\begin{equation}
\begin{aligned}
\label{barAC0}
  & \bar{A}_{\Theta}=\frac{1}{{4{\pi }^{2}}}
 \iint_{[0,2\pi]^2}
  \left( A_{\Theta} \right)\text{d}l\text{d}{l}_{J},\\
  & \bar{B}_{\Theta}=\frac{1}{{4{\pi }^{2}}}
 \iint_{[0,2\pi]^2}
  \left( B_{\Theta} \right)\text{d}l\text{d}{l}_{J},\\
   & \bar{C}_{\Theta}=\frac{1}{{4{\pi }^{2}}}
  \iint_{[0,2\pi]^2}
  \left(C_{\Theta} \right)\text{d}l\text{d}{l}_{J}\\
 \end{aligned}
\end{equation}
are the averaged values of $A_{\Theta}$, $B_{\Theta}$ and $C_{\Theta}$ respectively. 

The stability of the periodic orbits of the asteroid is guaranteed in linear approximation if the sequential principal minor $\bar{A}_{\Theta}>0$ and 
\begin{equation}
D_{\Theta}=\det\left[\begin{array}{cc}
\bar{A}_{\Theta} & \bar{B}_{\Theta} 
\\
\bar{B}_{\Theta} & \bar{C}_{\Theta} 
\end{array}\right] >0.
\end{equation}
The stability of the periodic orbits of the asteroid will be proved analytically with some conditions and then shown numerically for the general case with the help of Matlab.

\section{Limiting cases}

When the distance between the asteroid and the star is much smaller than between the planet and the star, an expansion over the ratio between $a$ and $a_J$ can be performed. This is called the inner case of the double-averaged spatial ER3BP. \citet{HLSY} considered the double-averaged value of the coefficients in the quadratic form of  $p_{3}$ and $q_{3}$ and averaged $W_{\Theta}$ with the same procedure in (\ref{barAC0}). The sequential principal minor 
\begin{equation}
\begin{aligned}
%\label{ABCtheta0}   
\bar{A}_{\Theta,\,\text{inner}} &\approx \frac{3 \left(1+4 e^{2}-5 e^{2} \cos \left(\Theta \right)^{2} \right)}{4 G \left(1-e_{J}^{2}\right)^{\frac{3}{2}}}\,a^{2}, 
\\
D_{\Theta,\,\text{inner}} &\approx  \frac{9 \left(1+3 e^{2}-4 e^{4}\right)}{16\, G^{2} \left(1-e_{J}^{2}\right)^{3}}\,a^{4},
 \end{aligned}
\end{equation}
are calculated to be positive for $a$ is small and $e_{J} \nrightarrow 1$. The orbits of the asteroid in the double-averaged model are stable in linear approximation for the inner case.

On the contrary, when the distance between the asteroid and the star is much larger than the distance between the planet and the star (outer case), an expansion of $d=1 / a$ can be performed analytically. Taking
\begin{equation}
M_{1}={\frac {x_{J}^{2}+y_{J}^{2}}{{x}^{2}+{y}^{2}+{z}^{2}}}, \quad  M_{2}=\,-\frac {2\left(xx_{J}+yy_{J}\right)}{{x}^{2}+{y}^{2}+{z}^{2}}, 
\end{equation}
and
\begin{equation}
M=M_{1}+M_{2}\,, 
\end{equation}
the approximate formula of the force function of gravity $U=U_{\Theta,\,\text{outer}}$ of the planet up to $a^{3}$ is expanded as
\begin{equation}
\label{V_expanded}
U_{\Theta,\,\text{outer}}=-\frac {1}{\sqrt {{x}^{2}+{y}^{2}+{z}^{2}}} \left(1-\frac{1}{2}M+\frac {3}{8} \left(M_{1} M_{2}+ M_{2}^{2}\right)-\frac{5}{16} M_{2}^{3}\right).
\end{equation}
Then, the double-averaged value of the coefficients in the quadratic form of  $p_{3}$ and $q_{3}$ is 
\begin{equation}
\label{W_out}
\bar{W}_{\Theta,\,\text{outer}}=\frac{1}{2}\left(\bar{A}_{\Theta,\,\text{outer}}\,{{p}_{3}}^{2}+2\bar{B}_{\Theta,\,\text{outer}} p_3 q_3+\bar{C}_{\Theta,\,\text{outer}}\, {{q}_{3}}^{2}\right).
\end{equation}

Averaging over mean anomalies is difficult. By Kepler's equation 
%\cite{scha}
\begin{equation}
l=E-e\sin E, \,\,\, l_J=E_J-e\sin E_J,
\end{equation}
the double-averaged value of $\bar{W}_{\Theta,\,\text{outer}}$ can be derived by
\begin{equation}
\label{averageway}
\begin{aligned}
   \bar{W}_{\Theta,\,\text{outer}}&=\frac{1}{{4{\pi }^{2}}} \iint_{[0,2\pi]^2}{{W}_{\Theta,\,\text{outer}}}\text{d}l\text{d}{{l}_{J}}
   \\
   &=\frac{1}{{4{\pi }^{2}}} \iint_{[0,2\pi]^2}{{W}_{\Theta,\,\text{outer}}}  
   \frac{\text{d}{{l}}}{\text{d}{{E}}}\frac{\text{d}{{l}_{J}}}{\text{d}{{E}_{J}}}{\text{d}E}\text{d}{{E}_{J}} 
   \\ 
 & =\frac{1}{{4{\pi }^{2}}} \iint_{[0,2\pi]^2}{{W}_{\Theta,\,\text{outer}}}\left( 1-e\cos  E  \right)\left( 1-{{e}_{J}}\cos  {{E}_{J}}  \right)\text{d}E\text{d}{{E}_{J}}. 
\end{aligned}
\end{equation}  
$\bar{A}_{\Theta,\,\text{outer}}$, $\bar{B}_{\Theta,\,\text{outer}}$ and $\bar{C}_{\Theta,\,\text{outer}}$ are derived similarly. Substituting $x$, $y$, $x_{J}$, $y_{J}$ into the double-averaged value of the coefficients, the coefficients up to order $d^4$ are
\begin{equation}
\label{ABCtheta0}   
\begin{aligned}
\bar{A}_{\Theta,\,\text{outer}}=&\frac{3 \left(1-e_{J}^{2}\right)}{4 \left(1-e^{2}\right)^{\frac{3}{2}} G}\,d^{3} -\frac{75\,e\,e_{J} \cos\left(\Theta \right) \left(1-e_{J}^{2}\right)}{32\left(1-e^{2}\right)^{\frac{5}{2}} G}\,d^{4} + O(d^5), 
\\ 
\bar{B}_{\Theta,\,\text{outer}}= &\frac{15\,e\,e_{J} \sin\left(\Theta \right) (17 e_{J}^{2}-24)}{128 (1-e^{2})^{\frac{5}{2}} G}\,d^{4} + O(d^5), 
\\ 
\bar{C}_{\Theta,\,\text{outer}}=&\frac{3 \left(1+4 e_{J}^{2}\right)}{4 \left(1-e^{2}\right)^{\frac{3}{2}} G}\,d^{3}-\frac{15 \,e\,e_{J}\cos\left(\Theta \right) \left(43 e_{J}^{2}+34\right) }{64 \left(1-e^{2}\right)^{\frac{5}{2}} G}\,d^{4} + O(d^5). 
\end{aligned}
\end{equation}

In the above formulas, as $d$ is small enough, for $e \nrightarrow 1$, all the coefficients of $d^4$ in (\ref{ABCtheta0}) are bounded, thus terms of $d^4$, as well as higher order terms, are omitted. It is obvious that $\bar{A}_{\Theta,\,\text{outer}}>0$ and $\bar{C}_{\Theta,\,\text{outer}}>0$. The determinant of the quadratic form (\ref{W_out}) up to order $a^7$ is 
\begin{equation}
\label{DT0}
D_{\Theta,\,\text{outer}}=\frac{9 \left(1-e_{J}^{2}\right) \left(1+4 e_{J}^{2}\right)}{16 \left(1-e^{2}\right)^{3} G^{2}}\,d^{6}+\frac{45 e \left(44 e_{J}+39 e_{J}^{3}-83 e_{J}^{5}\right) \cos\left(\Theta \right)}{256 \left(1-e^{2}\right)^{4} G^{2}}\,d^{7}
\end{equation}

Similarly, when $d$ is small, the coefficient of $d^7$ in (\ref{DT0}) is bounded as far as $e \nrightarrow 1$. For small values of $d$, the sequential principal minor $\bar{A}_{\Theta,\,\text{outer}}>0$ and $D_{\Theta,\,\text{outer}}>0$, thus (\ref{W_out}) is a positive definite quadratic form. Consequently, for the outer case, periodic orbits of the double-averaged planar elliptic restricted 3-body problem are stable in the linear approximation with spatial perturbation. 

\section{General case}

When the ratio between $a$ and $a_J=1$ is arbitrary (general case), the orbits of the asteroid and the planet are assumed not in collision, i.e., 
\begin{equation}
\label{condition}
\begin{aligned}
a \left(1+e \right)&< 1-e_{J} \quad \text{for}\,\,0<a<1,
\\
a \left(1-e \right)&> 1+e_{J} \quad \text{for}\,\,a>1.
\end{aligned}
\end{equation}
For periodic orbits around the equilibria in \citet{vash_planar}, the orbits of the asteroid sway up and down periodically. When the sum of the argument of the pericenter and the ascending node of the asteroid is $\Theta=\omega+\Omega$, denote the distance between the asteroid and the planet by

\begin{equation}
d_{\Theta}^3 (l,l_J)={\left[ {\left( x-{{x}_{J}} \right)}^{2}+{{\left( y-{{y}_{J}} \right)}^{2}} \right]}^{{3}/{2}}.
\end{equation}
Then substituting (\ref{coord_transform2'}) into (\ref{ABCR0}), we get

%\begin{equation}
%\label{ABCR}   
%v
%A_{\Theta}&=\,\frac{\sin(\Theta) x'{{y}_{J}}+\cos(\Theta) y'{{y}_{J}}}{G {\left[ {\left( x-{{x}_{J}} \right)}^{2}+{{\left( y-{{y}_{J}} \right)}^{2}} \right]}^{{3}/{2}}}, 
% \\ 
%B_{\Theta}&=\,\frac{\sin(\Theta) x'{{x}_{J}}+\cos(\Theta) y'{{x}_{J}}+\cos(\Theta) x'{{y}_{J}}-\sin(\Theta) y'{{y}_{J}}}{2G {{\left[ {{\left( x-{{x}_{J}} \right)}^{2}}+{{\left( y-{{y}_{J}} \right)}^{2}} \right]}^{{3}/{2}}}}, 
%\\
%C_{\Theta}&=\, \frac{\cos(\Theta) x'{{x}_{J}}-\sin(\Theta) y'{{x}_{J}}}{ G {\left[ {{\left( x-{{x}_{J}} \right)}^{2}}+{{\left( y-{{y}_{J}} \right)}^{2}} \right]^{{3}/{2}}}}.
%v
%\end{equation}

\begin{equation}
\label{ABCd}   
\begin{aligned}
A_{\Theta}&=\,\frac{y{{y}_{J}}}{G \,d_{\Theta}^3(l,l_J)}=\,\frac{1}{G \,d_{\Theta}^3(l,l_J)}\left[\sin(\Theta) x'{{y}_{J}}+\cos(\Theta) y'{{y}_{J}}\right], 
 \\ 
B_{\Theta}&=\,\frac{\left( x{{y}_{J}}+y{{x}_{J}} \right)}{2G \,d_{\Theta}^3(l,l_J)}=\,\frac{1}{2G \,d_{\Theta}^3(l,l_J)}\left[ \begin{aligned}
&\sin(\Theta) x'{{x}_{J}}+\cos(\Theta) y'{{x}_{J}}\\&+\cos(\Theta) x'{{y}_{J}}-\sin(\Theta) y'{{y}_{J}}\end{aligned} \right], 
\\
C_{\Theta}&=\, \frac{x{{x}_{J}}}{ G \,d_{\Theta}^3(l,l_J)}=\, \frac{1}{ G \,d_{\Theta}^3(l,l_J)}\left[\cos(\Theta) x'{{x}_{J}}-\sin(\Theta) y'{{x}_{J}}    
  \right].
\end{aligned}
\end{equation}
The double-averaged value of the quadratic form of  $p_{3}$ and $q_{3}$ is 
\begin{equation}
\label{bar{W}}
\bar{W}_{\Theta}=\frac{1}{2}\left(\bar{A}_{\Theta}\,{{p}_{3}}^{2}+2\bar{B}_{\Theta} p_3 q_3+\bar{C}_{\Theta}\, {{q}_{3}}^{2}\right),
\end{equation}
where 
\begin{equation}
\begin{aligned}
\label{barABC}
  & \bar{A}_{\Theta}=\frac{1}{{4{\pi}^{2} G}}
  \iint_{[0,2\pi]^2}
  \frac{1}{d_{\Theta}^3(l,l_J)}\left[\sin(\Theta) x'{{y}_{J}}+\cos(\Theta) y'{{y}_{J}}\right]\text{d}l\text{d}{l}_{J},\\
  & \bar{B}_{\Theta}=\frac{1}{{8{\pi }^{2} G}}
  \iint_{[0,2\pi]^2}
 \frac{1}{d_{\Theta}^3(l,l_J)}\left[ \begin{aligned}&\sin(\Theta) x'{{x}_{J}}+\cos(\Theta) y'{{x}_{J}}\\&+\cos(\Theta) x'{{y}_{J}}-\sin(\Theta) y'{{y}_{J}}\end{aligned} \right]\text{d}l\text{d}{l}_{J},\\
   & \bar{C}_{\Theta}=\frac{1}{{4{\pi }^{2} G}}
  \iint_{[0,2\pi]^2}
  \frac{1}{d_{\Theta}^3(l,l_J)}\left[\cos(\Theta) x'{{x}_{J}}-\sin(\Theta) y'{{x}_{J}} \right]\text{d}l\text{d}{l}_{J}\\
 \end{aligned}
\end{equation}

\subsection{General case with small $\Theta$}

In the real cosmos, the Lidov-Kozai effects mostly occur with small changes in eccentricities, thus it is important to consider the small periodic orbits around the equilibrium in \citet{vash_planar}. In this case, it is reasonable to assume that $\Theta$ is small, then
\begin{equation}
\begin{aligned}
d_{\Theta}^3 (l,l_J)&={\left[ {\left( x-{{x}_{J}} \right)}^{2}+{{\left( y-{{y}_{J}} \right)}^{2}} \right]}^{{3}/{2}}
\\
&={d'}_{\Theta}^3 (l,l_J)+3 d'_{\Theta} (l,l_J) \left( y' x_{J}- x' y_{J} \right) \Theta +O(\Theta^2),
 \end{aligned}
\end{equation}
where 
\begin{equation}
d'_{\Theta}(l,l_J)=\sqrt{\left( x'-{{x}_{J}} \right)^{2}+\left( y'-{{y}_{J}} \right)^{2}}.
\end{equation}
This subsection will prove the positive definiteness of the sequential principal minor of the quadratic form $\bar{W}_{\Theta}$ with small $\Theta$. The proof is under the assumption that the orbit of the asteroid is inside the planet's orbit. The outside case can be proved similarly. Firstly, let us show 
\begin{equation}
\label{x'yJ}
\iint_{[0,2\pi]^2}{\frac{x'{{y}_{J}}}{d_{\Theta}^3(l,l_J)}}\,dldl_J \sim O(\Theta),
\end{equation}
and 
\begin{equation}
\label{y'xJ}
\iint_{[0,2\pi]^2}{\frac{y'{{x}_{J}}}{d_{\Theta}^3(l,l_J)}}\,dldl_J \sim O(\Theta).
\end{equation}
\begin{figure}[htbp]
  \centering
\subfloat[]{\includegraphics[width=0.45\textwidth]{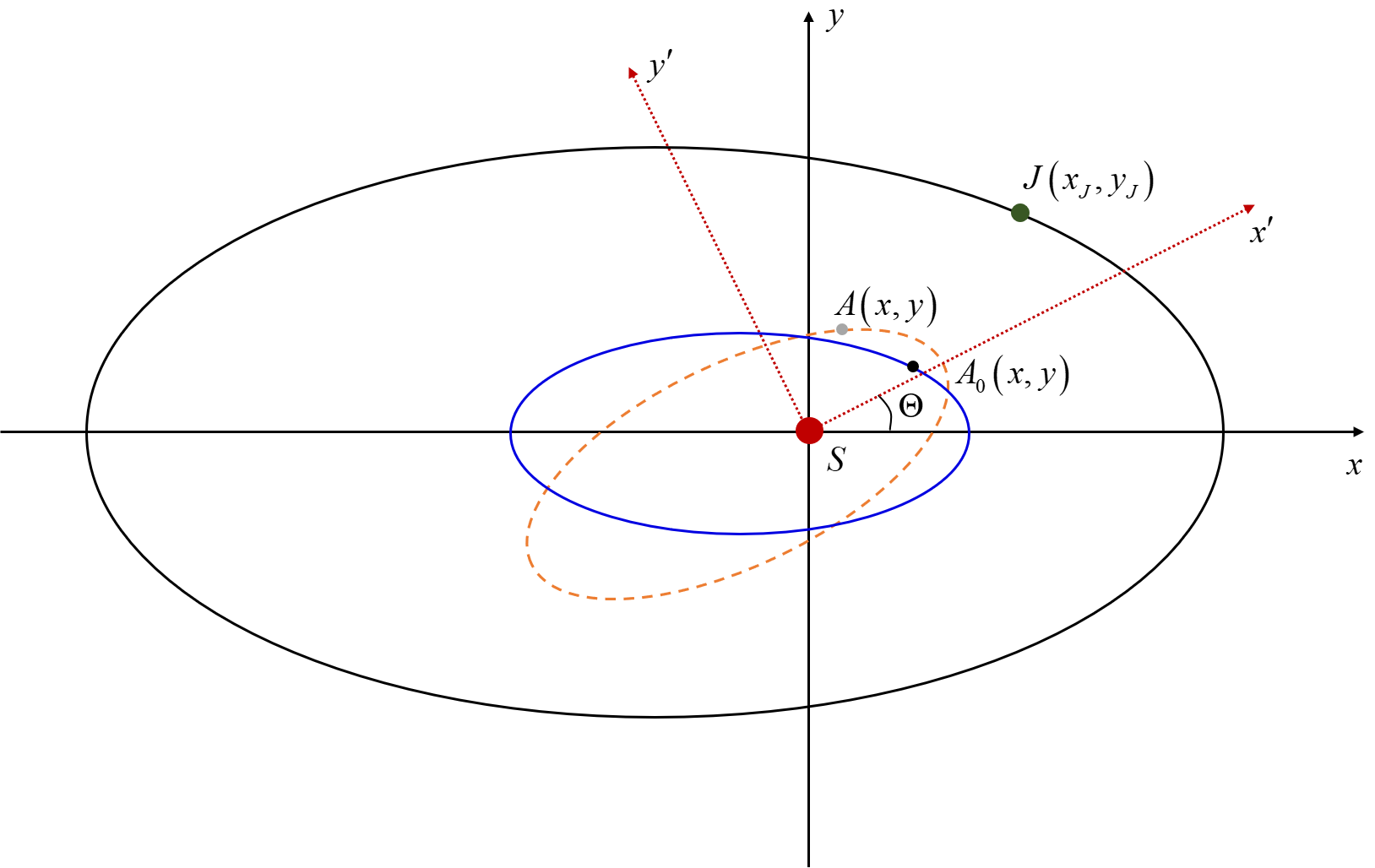}\label{ABCprove}} 
\subfloat[]{\includegraphics[width=0.45\textwidth]{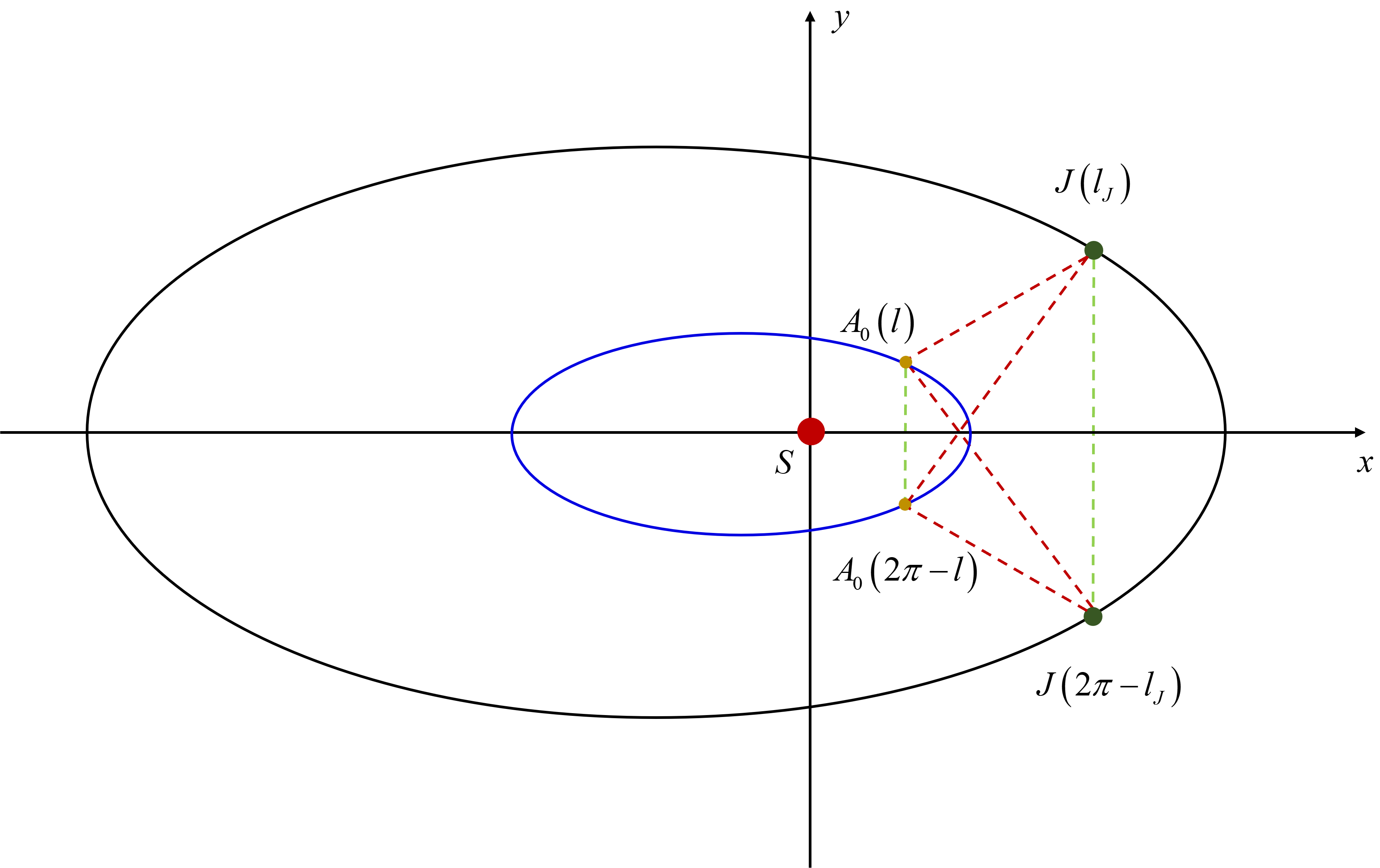}\label{Aprove}}
\caption{New orbit and symmetry of orbit.}
%\label{fig:Neworbit} 
\end{figure}

The orbit of the asteroid $A$ concerning $(x', y')$ can be regarded as an orbit of $A_0$ with $x=x'$ and $y=y'$, in detail, the blue orbit of $A_0$ can replace the orange orbit of $A$ in Fig. \ref{ABCprove}. It can be found that the new orbit (orbit of $A_0$) and the orbit of the planet are symmetrical about the $y$-axis, See Fig \ref{Aprove}, then
\begin{equation}
{d'}_{\Theta}(l,l_J)={d'}_{-\Theta}(2\pi-l,2\pi-l_J),\quad {d'}_{\Theta}(2\pi-l,l_J)={d'}_{-\Theta}(l,2\pi-l_J).
\end{equation}
Also, distances with expansions of $\Theta$ are
\begin{equation}
\begin{aligned}
&d_{\Theta}^3 (l,l_J)={d'}_{\Theta}^3 (l,l_J)+3 d'_{\Theta} (l,l_J) \left( y' x_{J}- x' y_{J} \right) \Theta ,
\\
&d_{\Theta}^3 (2\pi-l,2\pi-l_J)={d'}_{\Theta}^3 (l,l_J)- 3 d'_{\Theta} (l,l_J) \left( y' x_{J}- x' y_{J} \right) \Theta ,
\\
&d_{\Theta}^3 (l,2\pi-l_J)={d'}_{\Theta}^3 (l,2\pi-l_J)+3 d'_{\Theta} (l,2\pi-l_J) \left( y' x_{J}+ x' y_{J} \right) \Theta ,
\\
&d_{\Theta}^3 (2\pi-l,l_J)={d'}_{\Theta}^3 (l,2\pi-l_J)- 3 d'_{\Theta} (l,2\pi-l_J) \left( y' x_{J}+ x' y_{J} \right) \Theta .
\end{aligned}
\end{equation}
Then formula (\ref{x'yJ}) is proved by
\begin{equation*}
\begin{aligned}
&\iint_{[0,2\pi]^2}{\frac{x'{{y}_{J}}}{d_{\Theta}^3 (l,l_J)}}\,dldl_J
\\
=&\iint_{[0,\pi]^2} \left[\frac{x'{{y}_{J}}}{d_{\Theta}^3 (l,l_J)}-\frac{ x'{{y}_{J}}}{d_{\Theta}^3 (2\pi-l,2\pi- l_J)}+\frac{ x'{{y}_{J}}}{d_{\Theta}^3 (2\pi-l,l_J)}-\frac{ x'{{y}_{J}}}{d_{\Theta}^3 (l,2\pi-l_J)}       \right]\,
dldl_J
\\
=&\iint_{[0,\pi]^2} x'{{y}_{J}} \left[\frac{d_{\Theta}^3 (2\pi-l,2\pi- l_J)-d_{\Theta}^3 (l,l_J)}{d_{\Theta}^3 (l,l_J) d_{\Theta}^3 (2\pi-l,2\pi- l_J)}+\frac{d_{\Theta}^3 (l,2\pi-l_J)-d_{\Theta}^3 (2\pi-l,l_J)}{d_{\Theta}^3 (2\pi-l,l_J) d_{\Theta}^3 (l,2\pi-l_J)} \right]\,
dldl_J
\\
=& 6 \Theta \iint_{[0,\pi]^2} x'{{y}_{J}} \left[\frac{- d'_{\Theta} (l,l_J) \left( y' x_{J}- x' y_{J} \right)}{d_{\Theta}^3 (l,l_J) d_{\Theta}^3 (2\pi-l,2\pi- l_J)}+\frac{ d'_{\Theta} (l,2\pi-l_J) \left( y' x_{J}+ x' y_{J} \right) }{d_{\Theta}^3 (2\pi-l,l_J) d_{\Theta}^3 (l,2\pi-l_J)} \right]\,dldl_J 
\\
\sim & O(\Theta).
\end{aligned}
\end{equation*}
Similarly, 
\begin{equation*}
\begin{aligned}
&\iint_{[0,2\pi]^2}{\frac{y'{{x}_{J}}}{d_{\Theta}^3(l,l_J)}}\,dldl_J
\\
=&6 \Theta \iint_{[0,\pi]^2} y'{{x}_{J}} \left[\frac{- d'_{\Theta} (l,l_J) \left( y' x_{J}- x' y_{J} \right)}{d_{\Theta}^3 (l,l_J) d_{\Theta}^3 (2\pi-l,2\pi- l_J)}+\frac{ d'_{\Theta} (l,2\pi-l_J) \left( y' x_{J}+ x' y_{J} \right) }{d_{\Theta}^3 (2\pi-l,l_J) d_{\Theta}^3 (l,2\pi-l_J)} \right]\,dldl_J
\\
\sim & O(\Theta),
\end{aligned}
\end{equation*}
which proved formula (\ref{y'xJ}). Consequently, for small $\Theta$, 
\begin{equation}
\begin{aligned}
 \bar{B}_{\Theta}&=\frac{\Theta}{{8{\pi }^{2} G}} \iint_{[0,2\pi]^2}
 \frac{ x'{{x}_{J}}- y'{{y}_{J}}}{d_{\Theta}^3(l,l_J)}\text{d}l\text{d}{l}_{J} +\frac{\cos(\Theta)}{{8{\pi }^{2} G}} \iint_{[0,2\pi]^2}
 \frac{ y'{{x}_{J}}+ x'{{y}_{J}}}{d_{\Theta}^3(l,l_J)}\text{d}l\text{d}{l}_{J}
\\
&=\frac{\Theta}{{8{\pi }^{2} G}} \iint_{[0,2\pi]^2}
 \frac{ x'{{x}_{J}}- y'{{y}_{J}}}{d_{\Theta}^3(l,l_J)}\text{d}l\text{d}{l}_{J} +\frac{ O(\Theta)}{{8{\pi }^{2} G}} \sim O(\Theta).
 \end{aligned}
\end{equation}
Thus, $\bar{B}_{\Theta}$ is of order $\Theta$ in the case of small $\Theta$. Numerical checks of $\bar{B}_{\Theta}$ in Matlab for some values of $a$, $e_{J}$ are performed with variable $\Theta$ and $e$ taken from $0$ to $0.1$ and $0$ to $1$ respectively. In such a way we verified that $\bar{B}$ is small for $\Theta \in [0,0.1]$. 

\begin{figure}[htbp]
  \centering
{\includegraphics[width=0.45\textwidth]{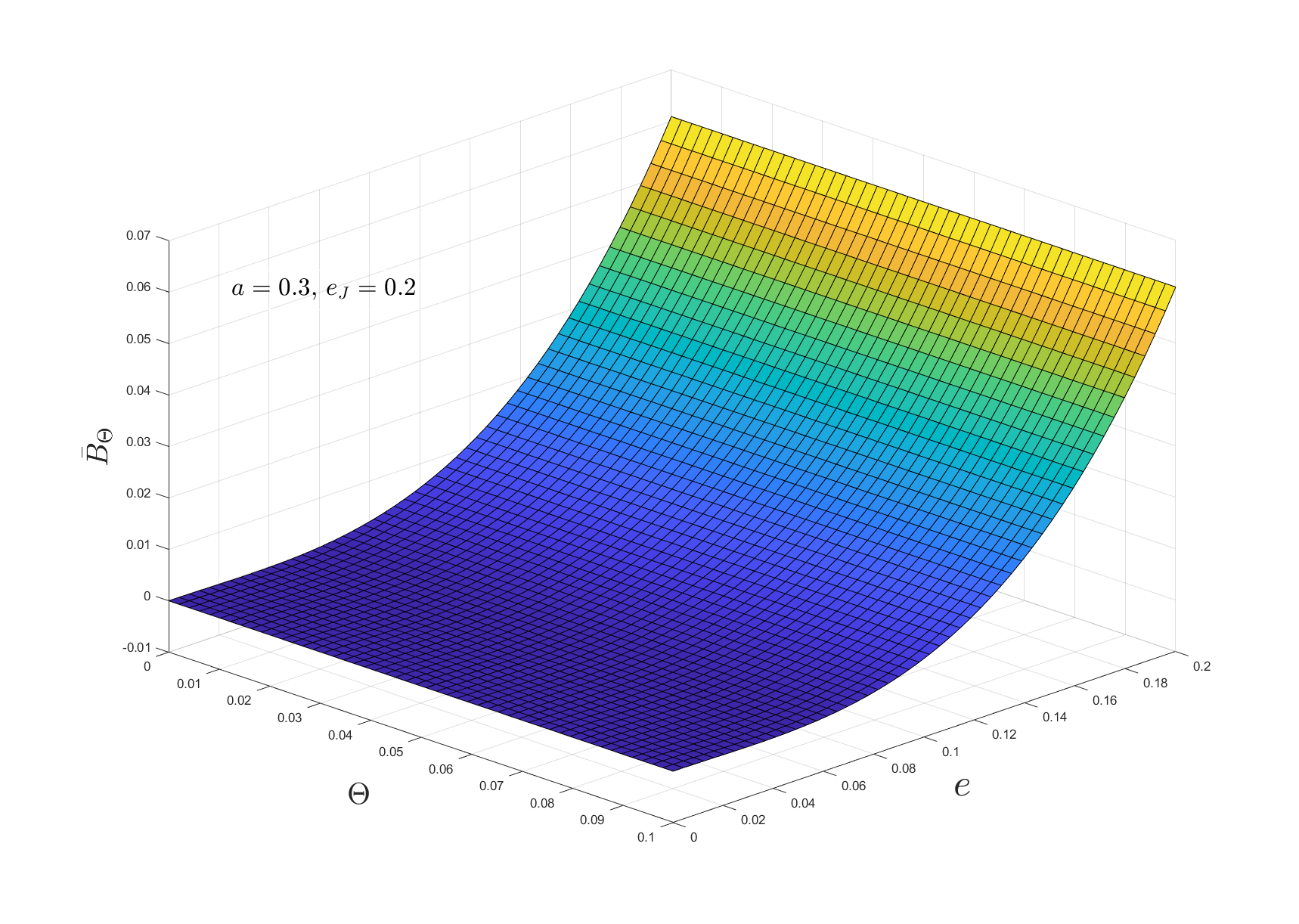}}
{\includegraphics[width=0.45\textwidth]{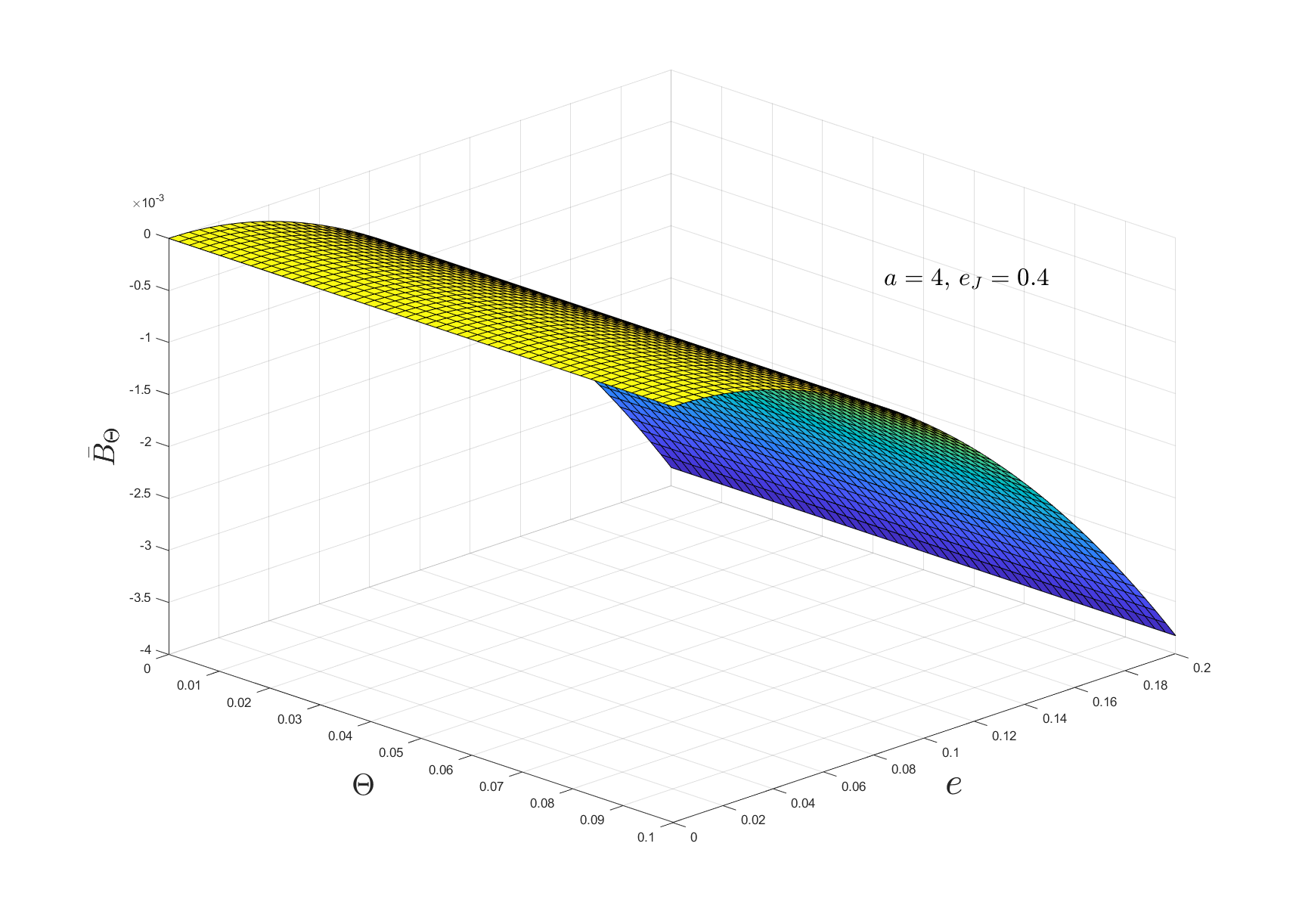}}
\caption{Value of $\bar{B}_{\Theta}$ for small $\Theta$.}
\label{fig:Bsmalltheta} 
\end{figure}

By formula (\ref{x'yJ}), the double-averaged value $\bar{A}_{\Theta}$ is
\begin{equation}
\label{barA}
\begin{aligned}
\bar{A}_{\Theta}&=\frac{\cos(\Theta)}{{4{\pi}^{2} G}}\iint_{[0,2\pi]^2}\frac{y'{{y}_{J}}}{d_{\Theta}^3(l,l_J)}\text{d}l\text{d}{l}_{J}+\frac{O(\Theta^2)}{{4{\pi}^{2} G}}
\\
&=\frac{1}{{2{\pi }^{2}G}}\iint_{[0,\pi]^2} \left[ \frac{d_{\Theta}^3(l,2\pi-l_J)-d_{\Theta}^3(l,l_J)}{d_{\Theta}^3(l,l_J) d_{\Theta}^3(l,2\pi-l_J)} \right]\,y' {y}_{J}\,dldl_J.
\end{aligned}
\end{equation}
For $l,l_J \in (0,\pi)$, it is obvious that 
\begin{equation}
 \frac{d_{\Theta}^3(l,2\pi-l_J)-d_{\Theta}^3(l,l_J)}{d_{\Theta}^3(l,l_J) d_{\Theta}^3(l,2\pi-l_J)} >0. 
\end{equation}
Thus $\bar{A}_{\Theta}$ is positive. Numerical results of $A$ in Matlab with $a=0.3$, $e_{J}=0.2$ and $a=4$, $e_{J}=0.4$ coincide with the analytical conclusion. 

\begin{figure}[htbp]
  \centering
{\includegraphics[width=0.45\textwidth]{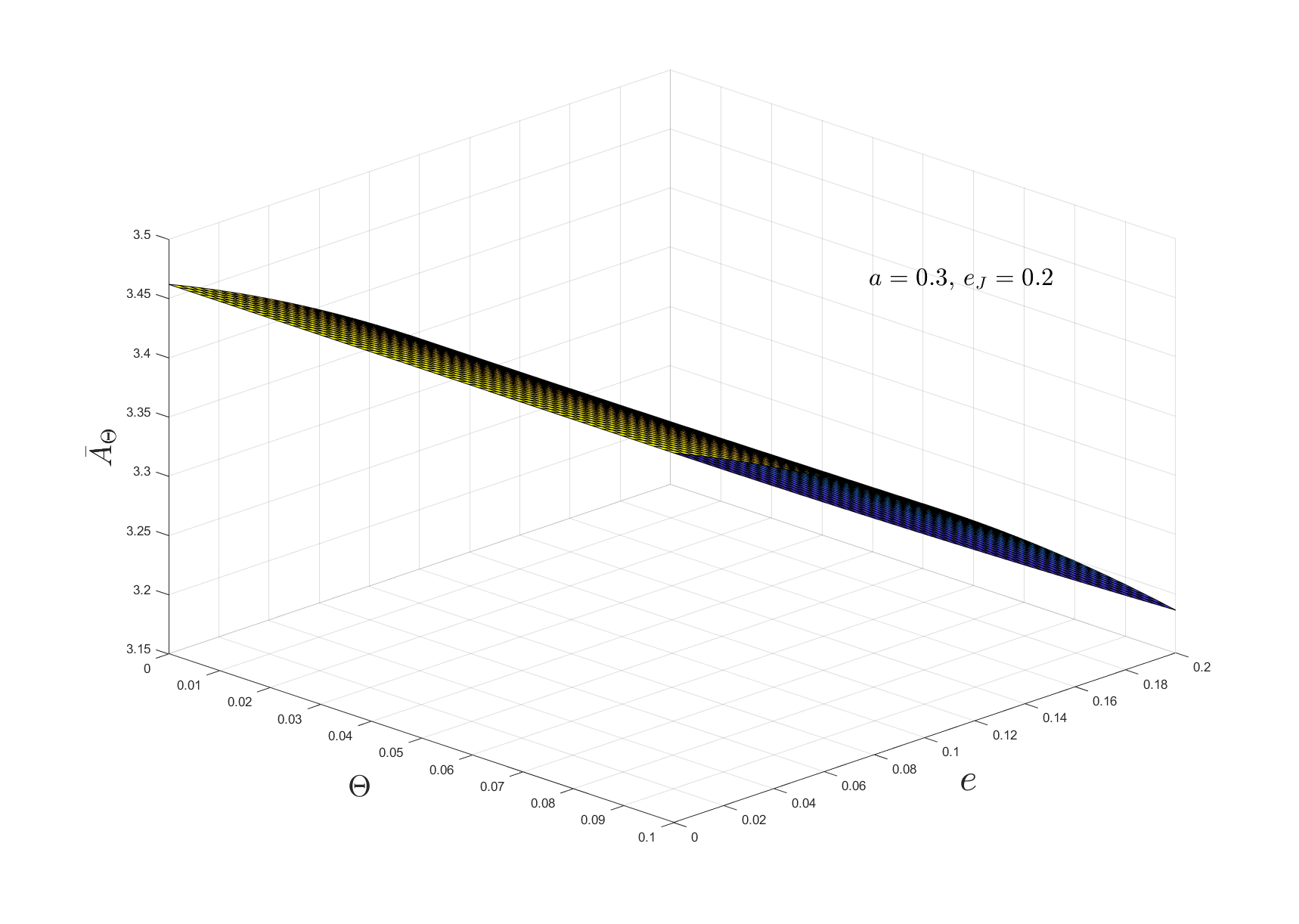}}
{\includegraphics[width=0.45\textwidth]{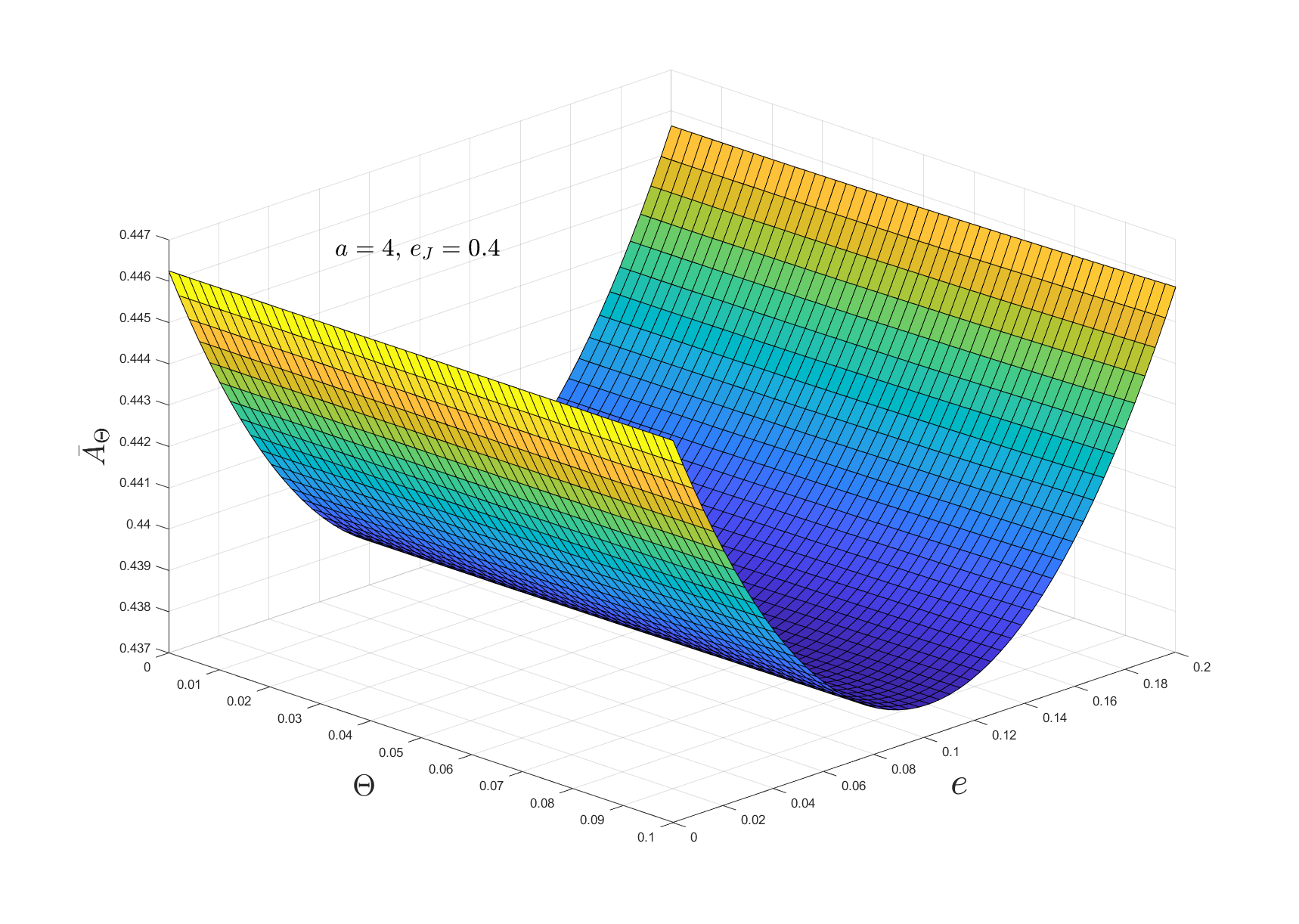}}
\caption{Value of $\bar{A}_{\Theta}$ for small $\Theta$.}
\label{fig:Asmalltheta} 
\end{figure}

By formula (\ref{y'xJ}), the double-averaged value $\bar{C}_{\Theta}$ is
\begin{equation}
\label{barC}
\begin{aligned}
\bar{C}_{\Theta}&=\frac{\cos(\Theta)}{{4{\pi}^{2} G}}\iint_{[0,2\pi]^2}\frac{x'{{x}_{J}}}{d_{\Theta}^3(l,l_J)}\text{d}l\text{d}{l}_{J}+\frac{O(\Theta^2)}{{4{\pi}^{2} G}}
\\
&=\frac{1}{{2{\pi }^{2}G}}\iint_{[0,\pi]^2} \frac{x'{{x}_{J}}}{d_{\Theta}^3(l,l_J)}\,dldl_J.
\end{aligned}
\end{equation}
Therefore, proving formula (\ref{barC}) is positive is equivalent to proving
\begin{equation}
\label{barC0}
\iint_{[0,\pi]^2}\frac{x{{x}_{J}}}{d_{\Theta}^3(l,l_J)}\text{d}l\text{d}{l}_{J}>0, 
\end{equation}
where $x=x'$, i.e. the orbit of asteroid $A$ can be considered as an orbit of $A_0$ in apsidal alignment case. The case of $e_{J}=0$ is the circular problem proved in \citep{neish}, thus $e_{J}\neq 0$ can be assumed.
\begin{figure}[htbp]
  \centering
\subfloat[]{\includegraphics[width=0.45\textwidth]{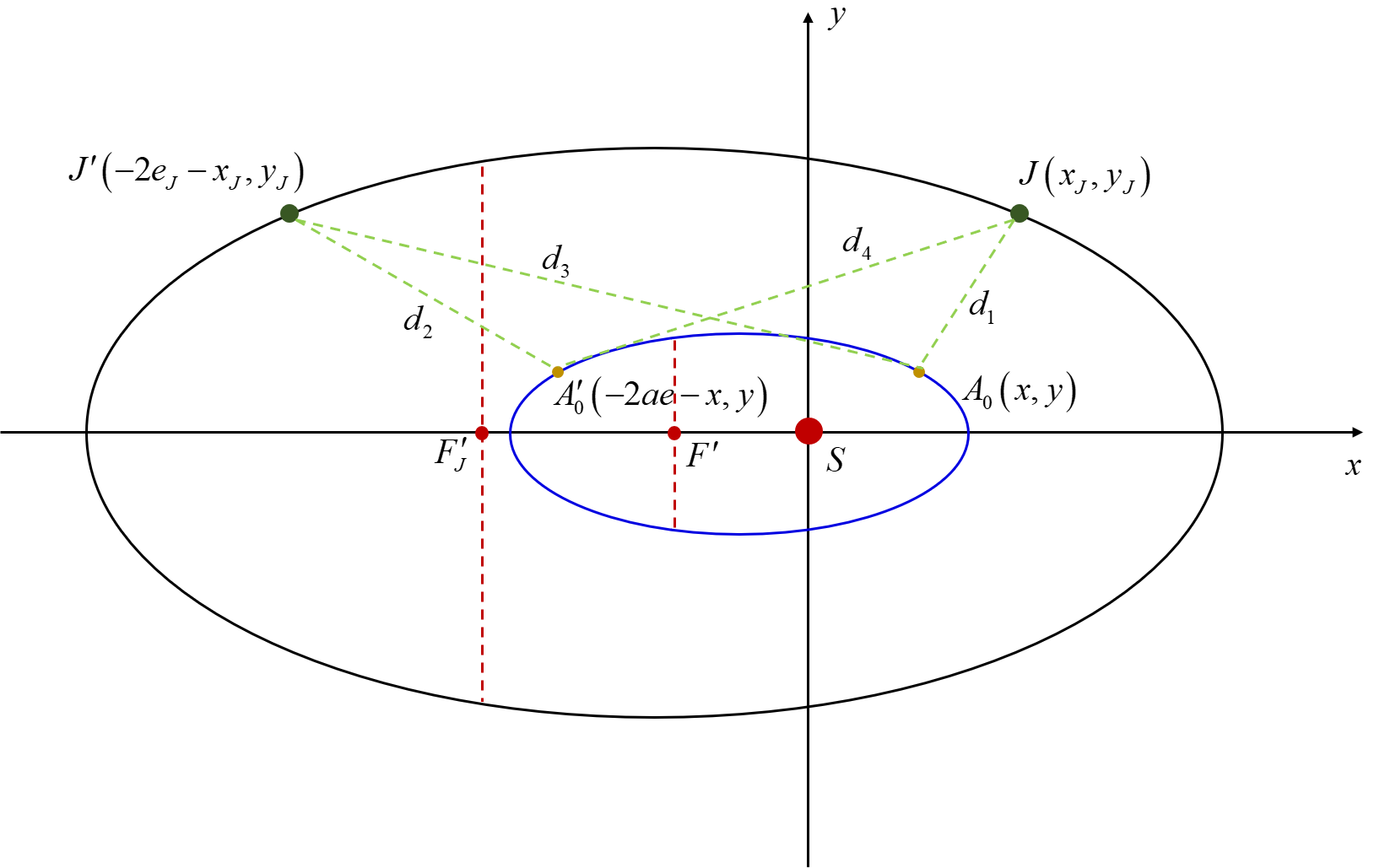}\label{Cprove1}} 
\subfloat[]{\includegraphics[width=0.45\textwidth]{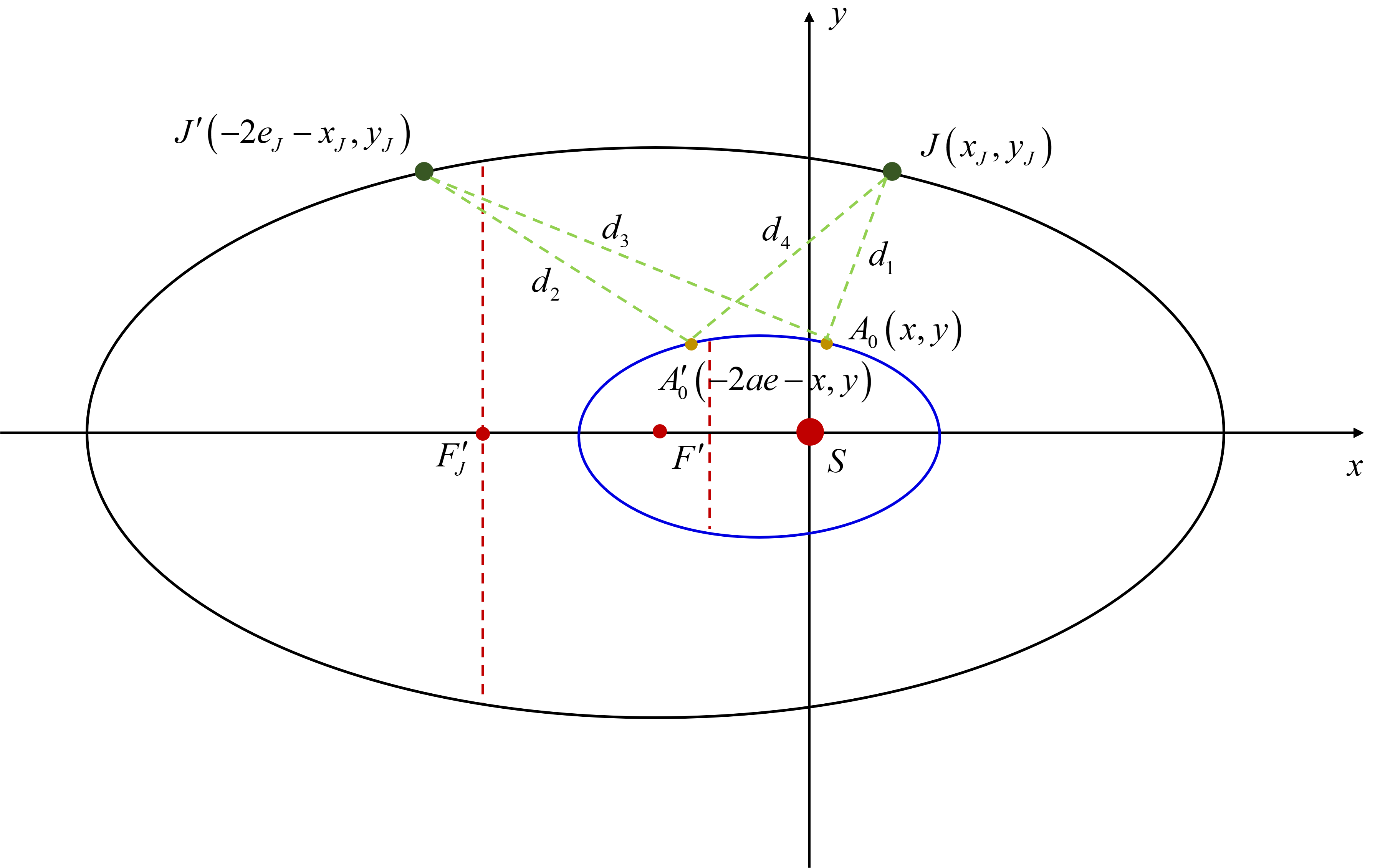}\label{Cprove2}} 
\caption{Prove of $C$.}
\label{fig:Cprove} 
\end{figure}
%When $x$ is taken between two focuses $F'$ and $S$ of the orbit of $A_0$, denoted by $A_0 \in I^{M}$, and $x_J$ is taken between two focuses $F'_J$ and $S$ of the orbit of $J$ ($J \in I_J^{M}$), we have $x<0$ and $x_J <0$. In this case,
%\begin{equation}
%\iint_{I^{M}\times I_J^{M}}\frac{x{{x}_{J}}}{d_{\Theta}^3(l,l_J)}\text{d}l\text{d}{l}_{J}>0
%\end{equation}
%Taking $x>0$, which means $A_0 \left(x,y \right)$ is on the right of $S$, denoted by $A_0 \in I^{R}$, then $A'_{0}\left(-2ae-x,y \right)$ is on the left of $F'$, denoted by $A'_{0} \in I^{L}$. Similarly, we take $J \left(x_{J},y_{J} \right)$ and $J' \left(-2e_{J}-x_{J},y_{J} \right)$ as well, denoted by $J \in I_J^{R}$, $J' \in I_J^{L}$. See Fig. \ref{fig:Cprove}. The distance $\overline{A_{0}J}=d_1$, $\overline{A'_{0} J'}=d_2$, $\overline{A_{0} J'}=d_3$ and $\overline{A'_{0} J}=d_4$. 

Taking $x>-ae$, which means $A_0 \left(x,y \right)$ is on the right of the orbit of the asteroid, denoted by $A_0 \in I^{R}$, then $A'_{0}\left(-2ae-x,y \right)$ is on the left, denoted by $A'_{0} \in I^{L}$. Similarly, taking $x>e_{J}$, then $J \left(x_{J},y_{J} \right) \in I_J^{R}$ and $J' \left(-2e_{J}-x_{J},y_{J} \right) \in I_J^{L}$. See Fig. \ref{fig:Cprove}. The distance $\overline{A_{0}J}=d_1$, $\overline{A'_{0} J'}=d_2$, $\overline{A_{0} J'}=d_3$ and $\overline{A'_{0} J}=d_4$. It is shown in Fig. \ref{Cprove1}, when $d_{2}<d_{4}$, $d_{2}<d_{3}$, $d_{1}<d_{4}$ and $d_{1}<d_{2}$, the considered function
\begin{equation}
\begin{aligned}
\hat{\mathcal{C}}=&\frac{x x_{J}}{d_1^3}+\frac{(-2ae-x)(-2e_{J}-x_{J})}{d_2^3}+\frac{x(-2e_{J}-x_{J})}{d_3^3}+\frac{(-2ae-x)x_{J}}{d_4^3}
\\
>&\frac{x x_{J}}{d_2^3}+\frac{(-2ae-x)(-2e_{J}-x_{J})}{d_2^3}+\frac{x(-2e_{J}-x_{J})}{d_2^3}+\frac{(-2ae-x)x_{J}}{d_2^3}
\\
=&\frac{4aee_{J}}{d_2^3}>0.
\end{aligned}
\end{equation}
If replacing $d_{2}<d_{4}$ by $d_{2}>d_{4}$, see Fig. \ref{Cprove2}, then by rearrangement inequality, 
\begin{equation}
\begin{aligned}
\hat{\mathcal{C}}\le&\frac{x x_{J}}{d_1^3}+\frac{(-2ae-x)(-2e_{J}-x_{J})}{d_4^3}+\frac{x(-2e_{J}-x_{J})}{d_3^3}+\frac{(-2ae-x)x_{J}}{d_2^3}
\\
>&\frac{x x_{J}}{d_4^3}+\frac{(-2ae-x)(-2e_{J}-x_{J})}{d_4^3}+\frac{x(-2e_{J}-x_{J})}{d_4^3}+\frac{(-2ae-x)x_{J}}{d_4^3}
\\
=&\frac{4aee_{J}}{d_4^3}>0.
\end{aligned}
\end{equation}
For other cases, one can prove them by rearrangement inequality directly. Thus, the integral
\begin{equation}
%\label{barC0}
\begin{aligned}
&\iint_{[0,\pi]^2}\frac{x{{x}_{J}}}{d_{\Theta}^3(l,l_J)}\text{d}l\text{d}{l}_{J}
\\
=&\left[\iint_{I^{L}\times I_J^{L}}+\iint_{I^{L}\times I_J^{R}}+\iint_{I^{R}\times I_J^{L}}+\iint_{I^{R}\times I_J^{R}} \right]\frac{x{{x}_{J}}}{d_{\Theta}^3(l,l_J)}\text{d}l\text{d}{l}_{J}
\\
=&\iint_{I^{R}\times I_J^{R}} \hat{\mathcal{C}} \text{d}l\text{d}{l}_{J}>0
%\iint_{I^{M}\times I_J^{M}}\frac{x{{x}_{J}}}{d_{\Theta}^3(l,l_J)}\text{d}l\text{d}{l}_{J}>0
\end{aligned}
\end{equation}

\begin{figure}[htbp]
  \centering
{\includegraphics[width=0.45\textwidth]{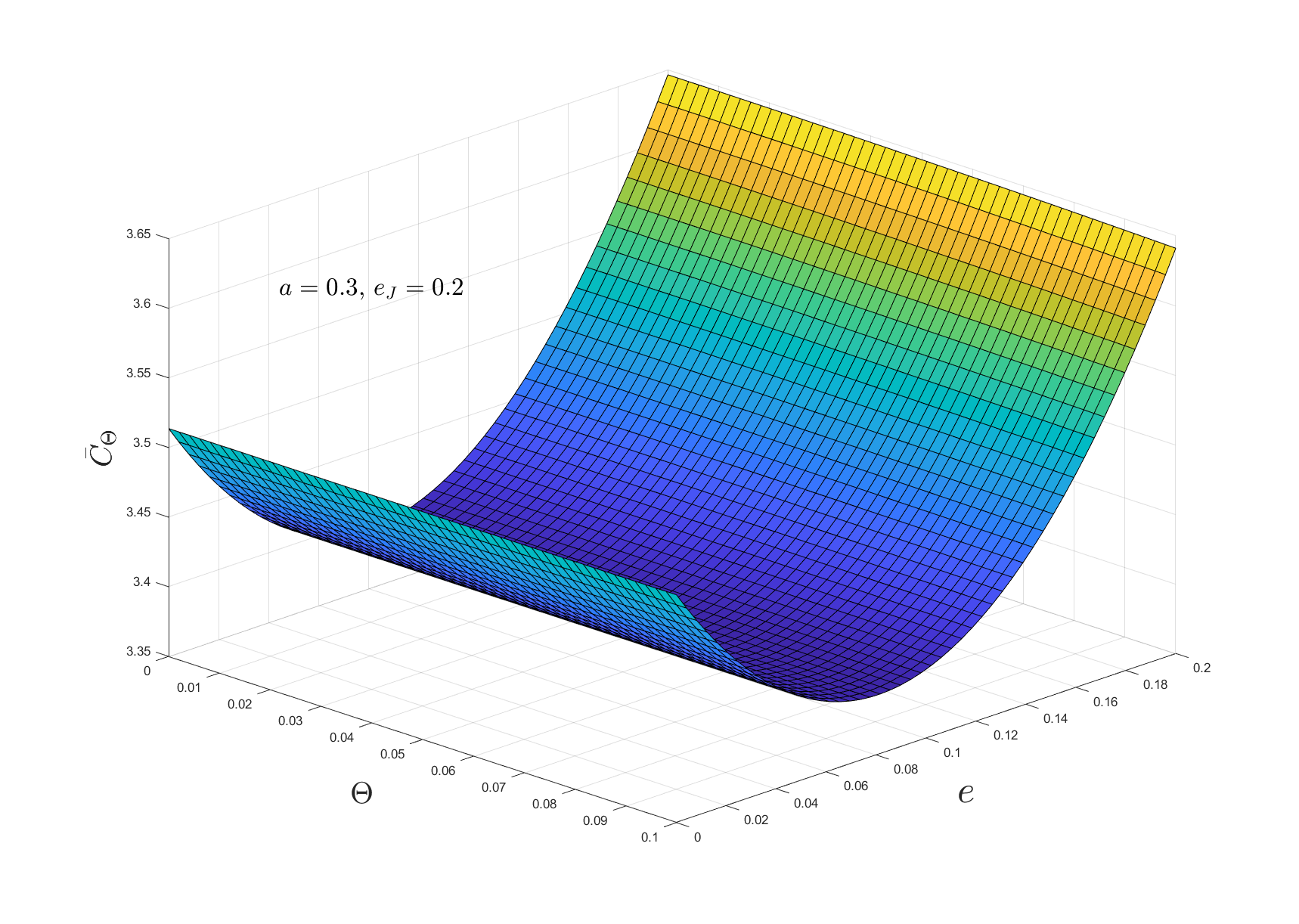}}
{\includegraphics[width=0.45\textwidth]{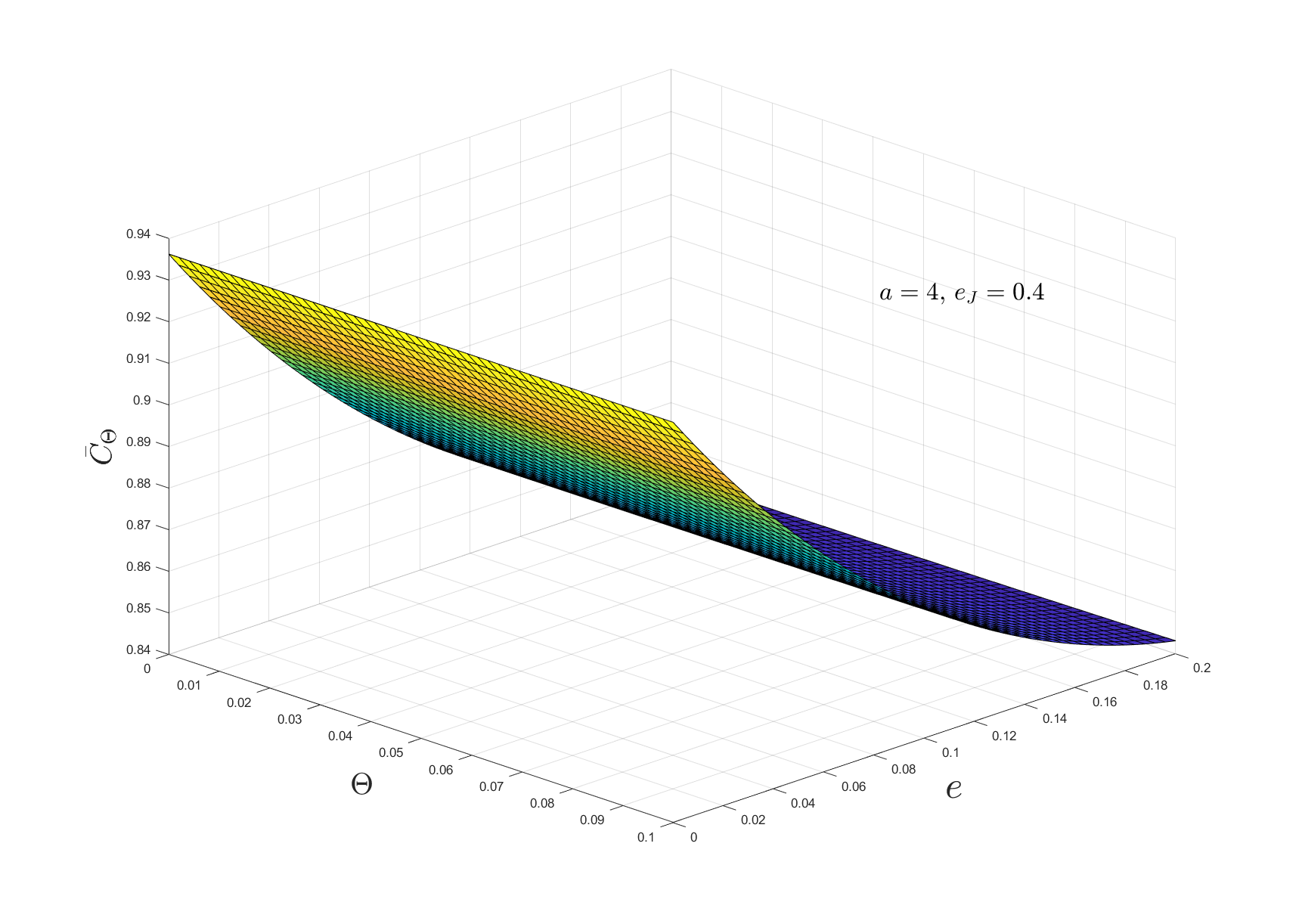}}
\caption{Value of $\bar{C}_{\Theta}$ for small $\Theta$.}
\label{fig:Csmalltheta} 
\end{figure}

As a result, $\bar{C}_{\Theta}$ is positive. Meanwhile, this is also an analytical proof of variable $C$ in \citet{NSS} instead of the numerical work. Numerical calculations of $\bar{C}_{\Theta}$ were performed in Matlab for some values of $a$, $e_{J}$. The variables $\Theta$ and $e$ are taken from $0$ to $0.1$ and $0$ to $1$ respectively with some grids. In such a way we verified that $\bar{C}$ is always positive. Fig. \ref{fig:Csmalltheta} are some numerical results with $a=0.3$, $e_{J}=0.2$ and  $a=4$, $e_{J}=0.4$.

%We calculated the value of $\bar C$ numerically. The function $\bar R$ is needed to find the value of the eccentricity of the asteroid in the equilibrium solution.

\begin{figure}[htbp]
  \centering
{\includegraphics[width=0.45\textwidth]{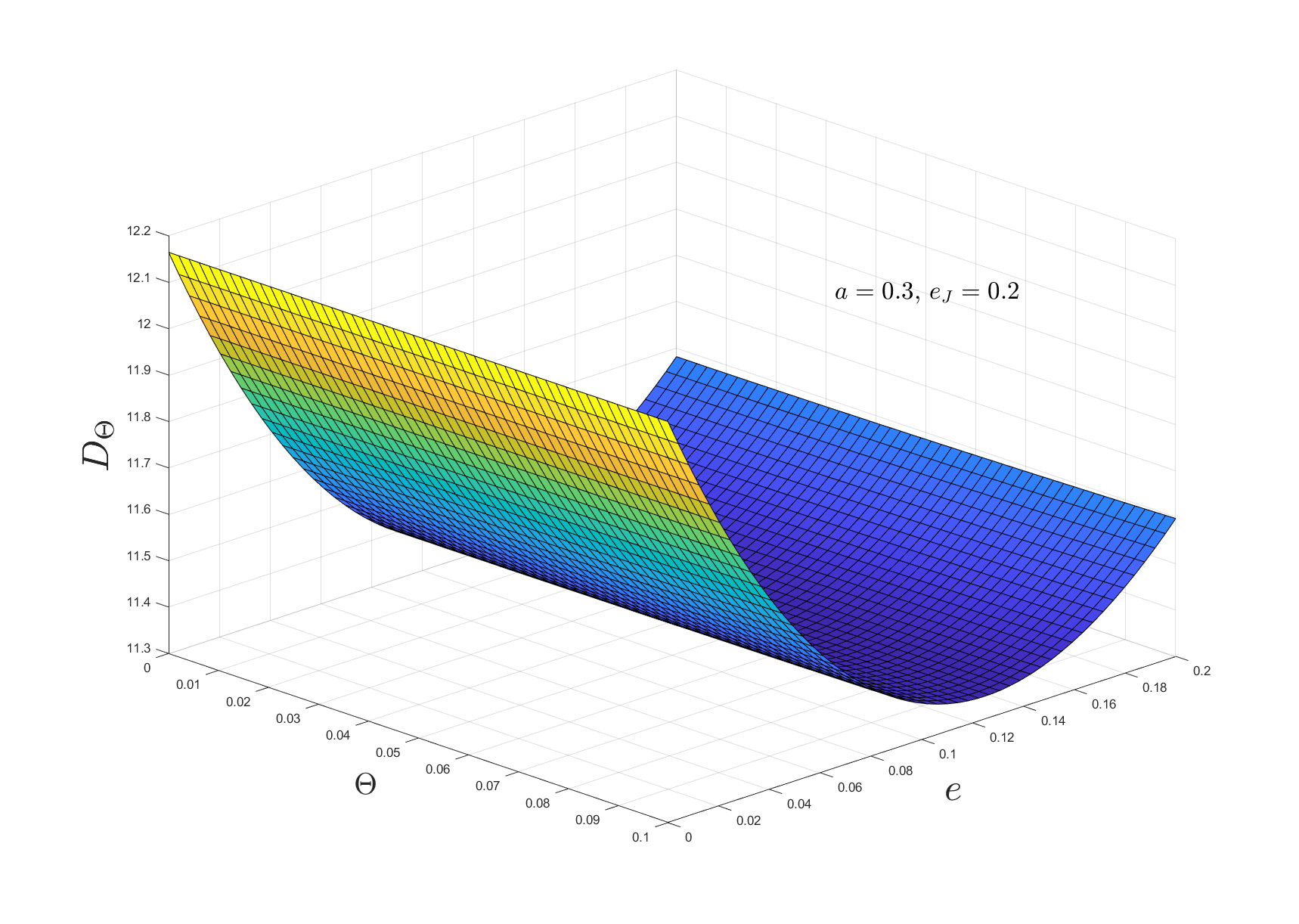}}
{\includegraphics[width=0.45\textwidth]{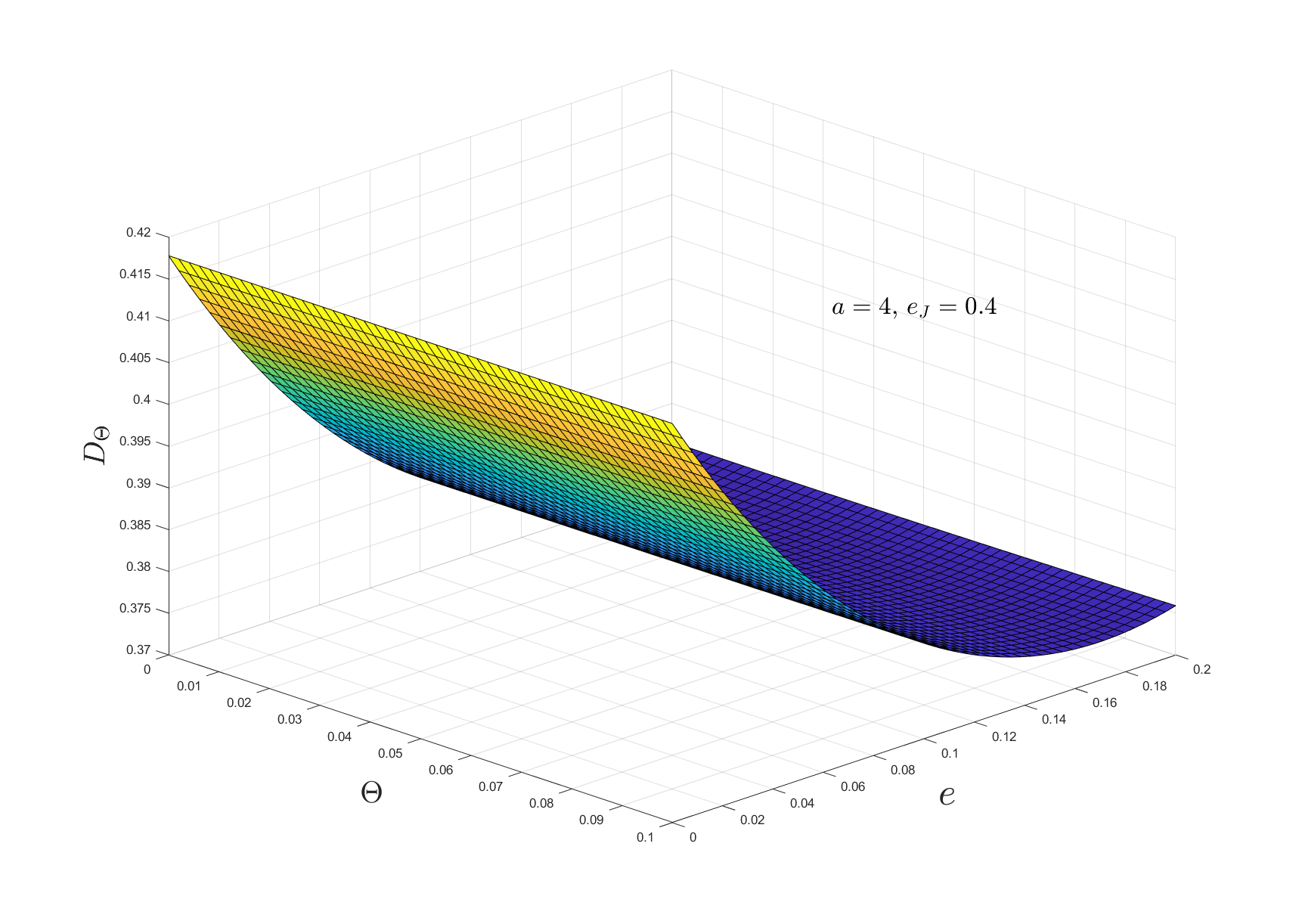}}
\caption{Value of $D_{\Theta}$ for small $\Theta$.}
\label{fig:Dsmalltheta} 
\end{figure}

Since $\bar{B}_{\Theta} \sim O(\Theta)$ is rather smaller than $\bar{A}_{\Theta}$ and $\bar{C}_{\Theta}$, the determinant $D_{\Theta}=\bar{A}_{\Theta} \bar{C}_{\Theta}-O(\Theta^2)>0$. Numerical checks are shown in Fig. \ref{fig:Dsmalltheta}. Thus, $\bar{A}_{\Theta}>0$,  $D_{\Theta}>0$, and  $\bar{W}_{\Theta}$ is a positive definite quadratic form. Hence, small periodic orbits around the equilibria of the double-averaged planar elliptic restricted 3-body problem are stable in the linear approximation as small periodic orbits of the double-averaged spatial perturbed elliptic restricted 3-body problem for small $\Theta$. 

\subsection{General case with arbitrary $\Theta$}

\begin{figure}[htbp]
  \centering
{\includegraphics[width=0.45\textwidth]{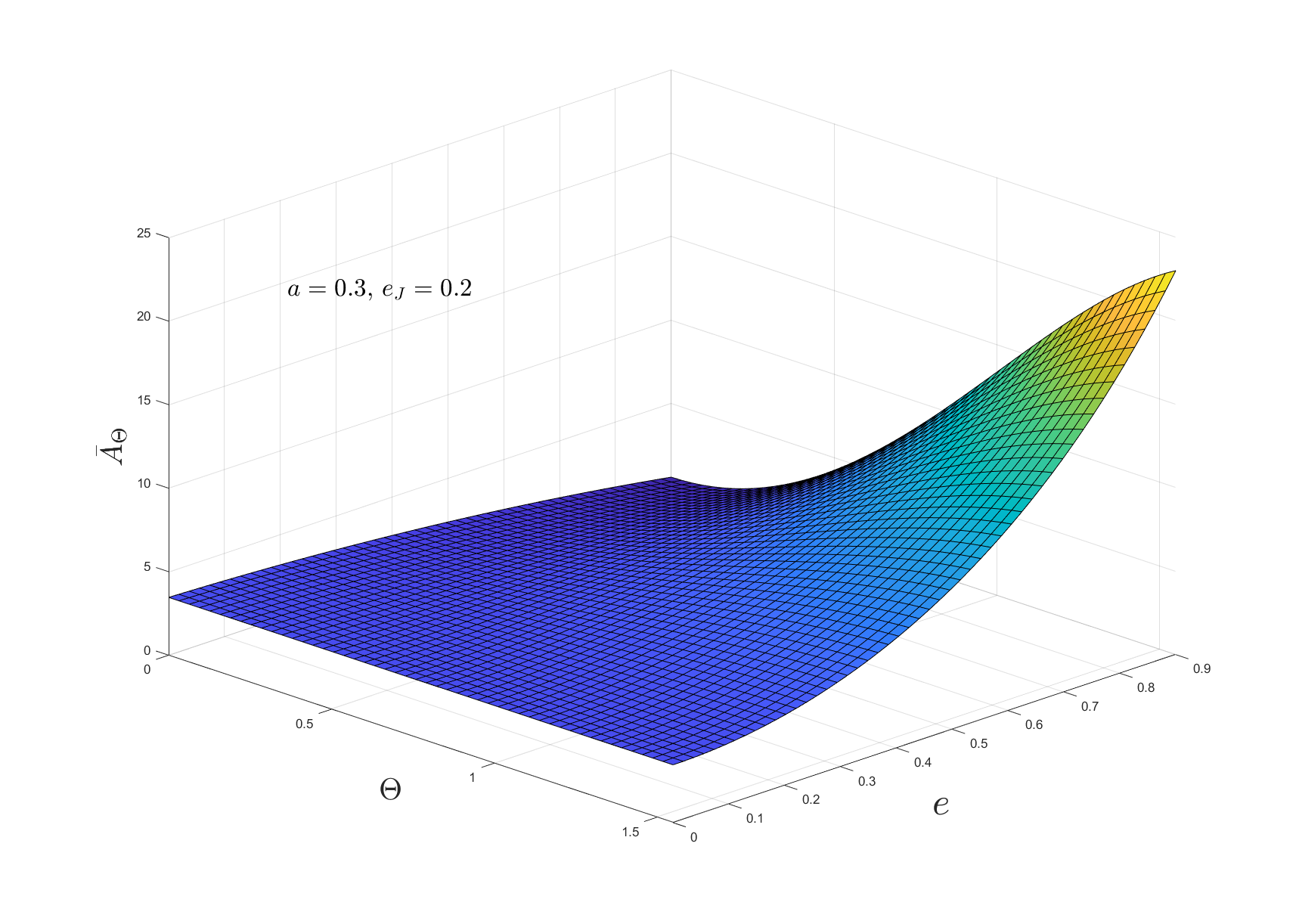}}{\includegraphics[width=0.45\textwidth]{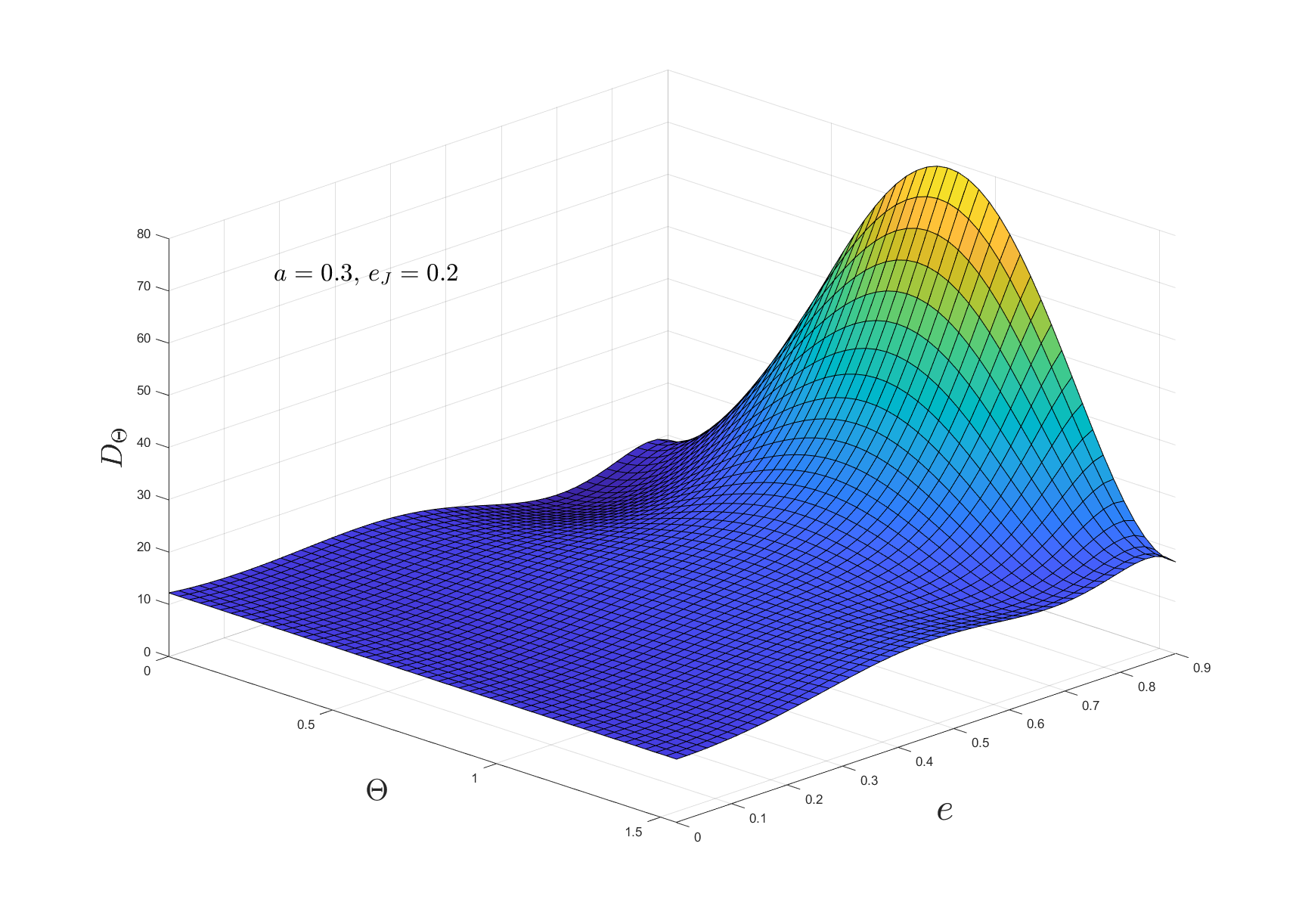}}
\caption{Value of $\bar{A}_{\Theta}$ and $D_{\Theta}$ for $a=0.3$ and $e_{J}=0.2$.}
\label{numeric1} 
\end{figure}

\begin{figure}[htbp]
  \centering
{\includegraphics[width=0.45\textwidth]{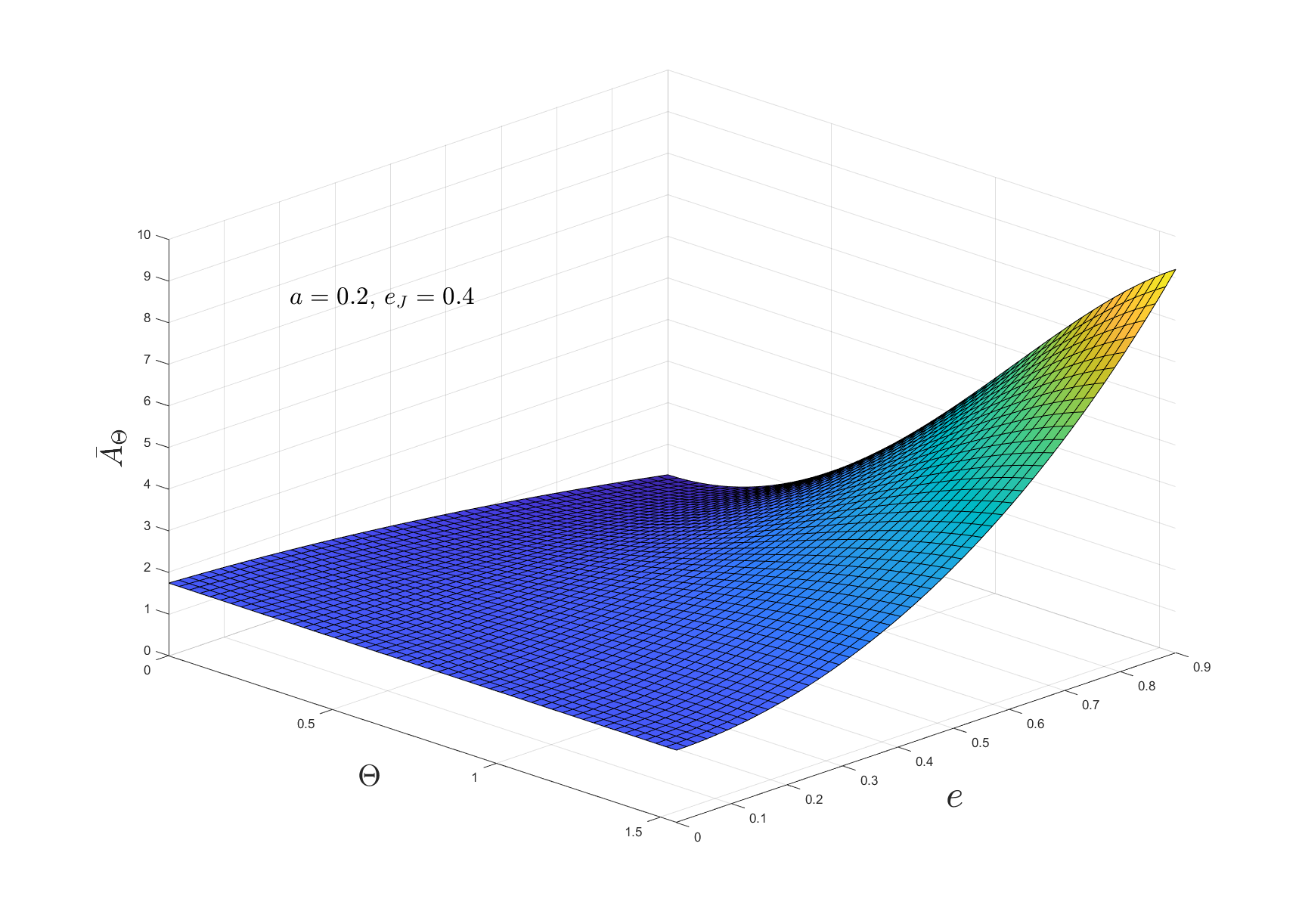}}{\includegraphics[width=0.45\textwidth]{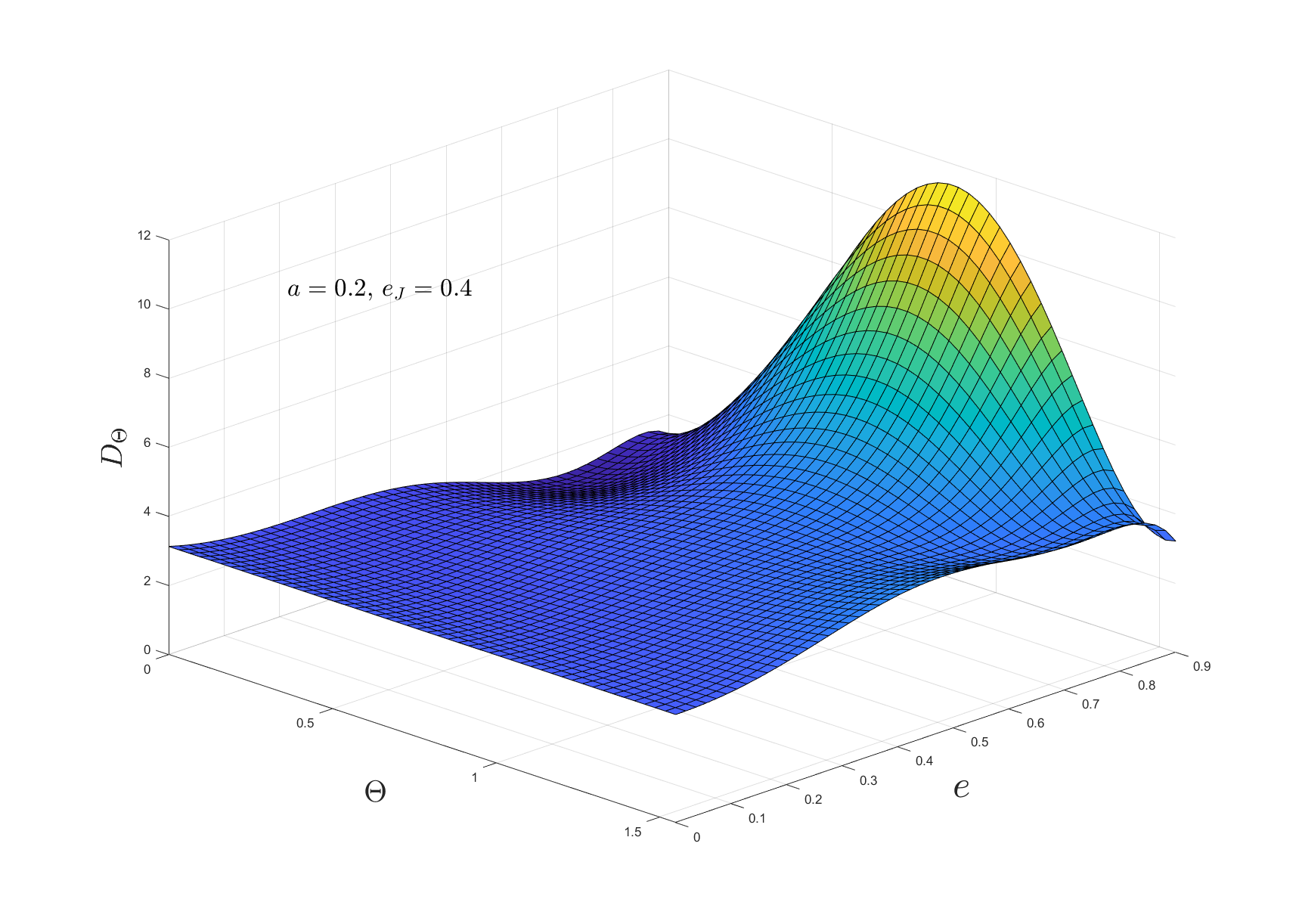}}
\caption{Value of $\bar{A}_{\Theta}$ and $D_{\Theta}$ for $a=0.2$ and $e_{J}=0.4$.}
\label{numeric2} 
\end{figure}

\begin{figure}[htbp]
  \centering
{\includegraphics[width=0.45\textwidth]{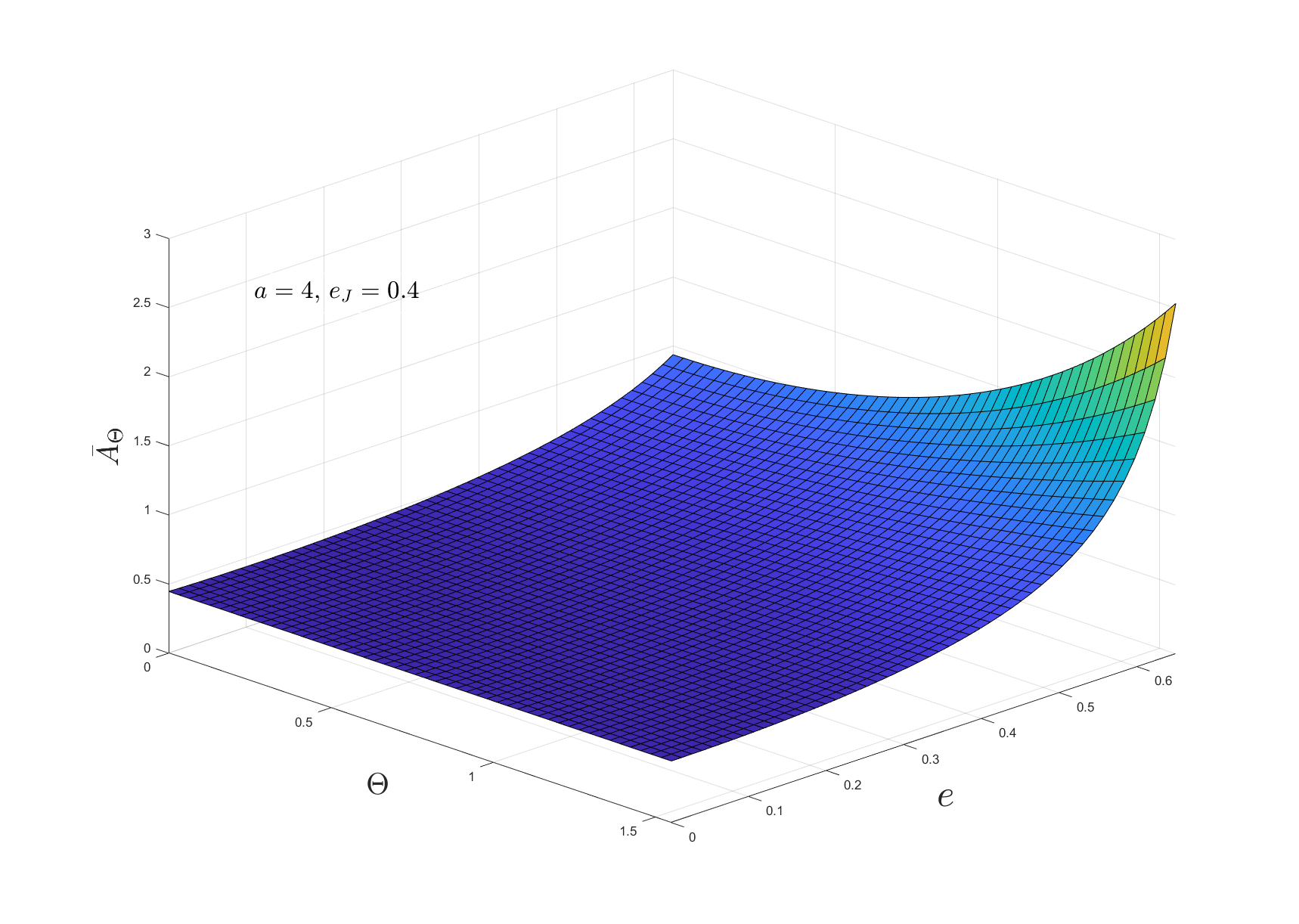}}{\includegraphics[width=0.45\textwidth]{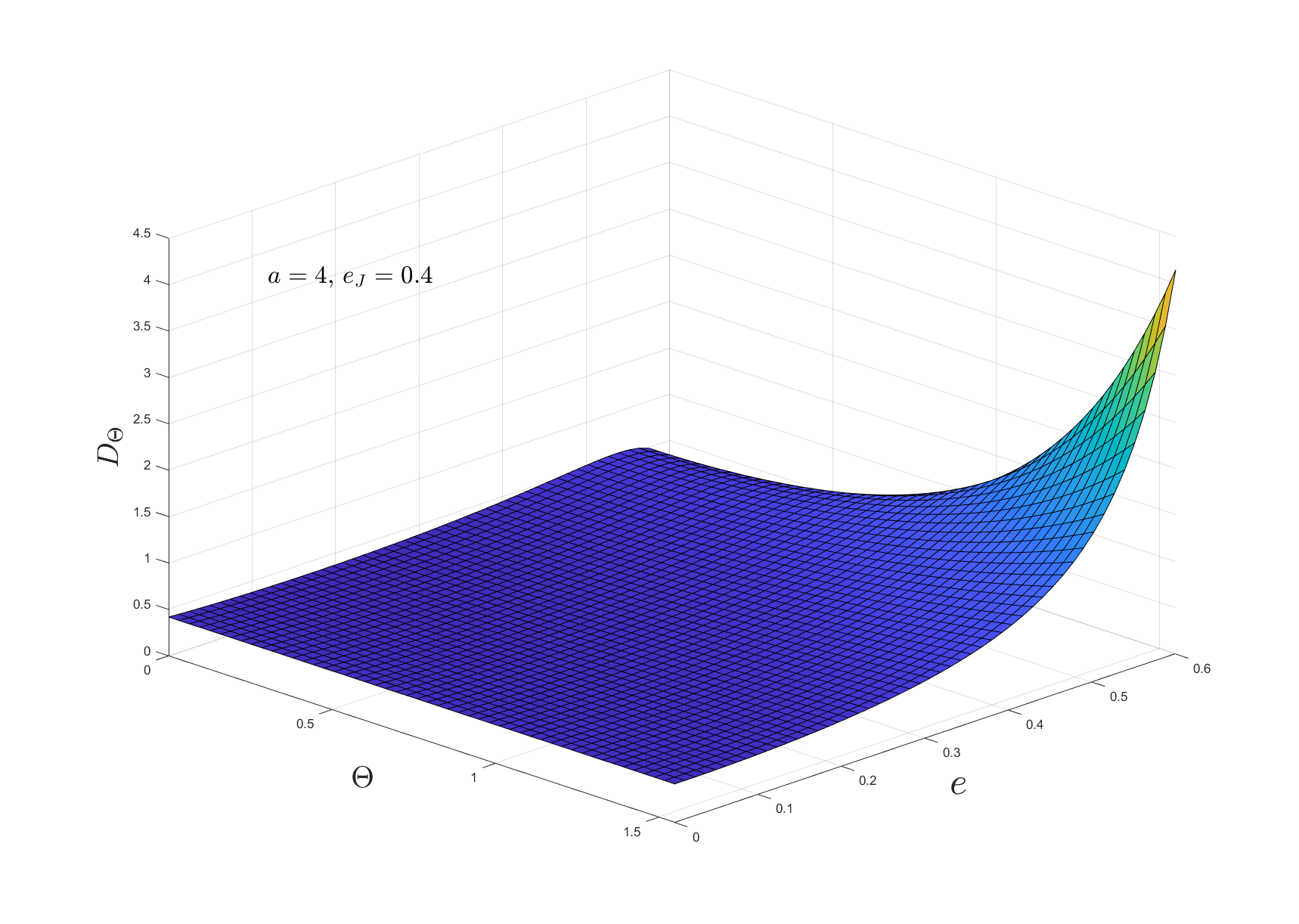}}
\caption{Value of $\bar{A}_{\Theta}$ and $D_{\Theta}$ for $a=4$ and $e_{J}=0.4$.}
\label{numeric3} 
\end{figure}

\begin{figure}[htbp]
  \centering
{\includegraphics[width=0.45\textwidth]{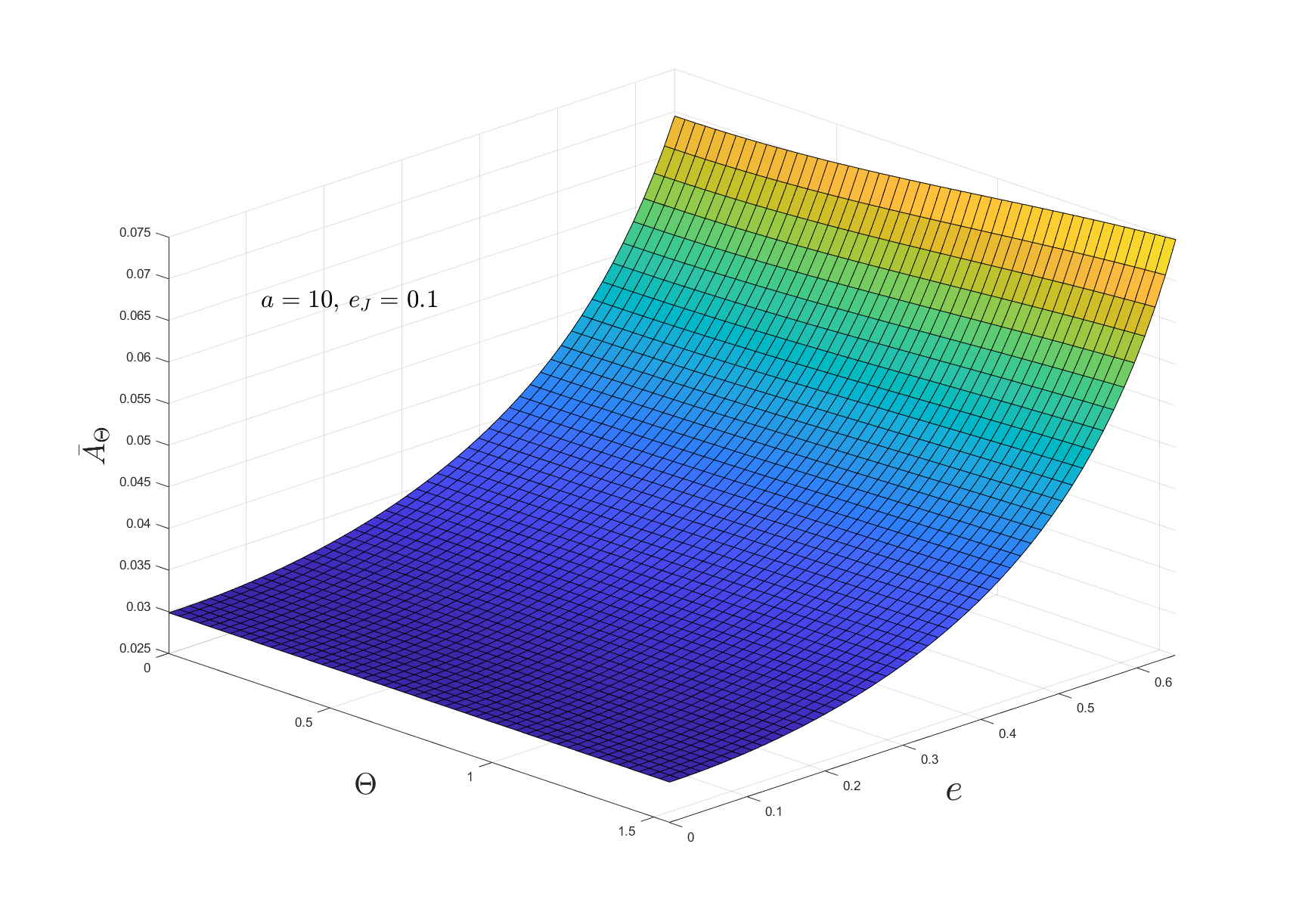}}{\includegraphics[width=0.45\textwidth]{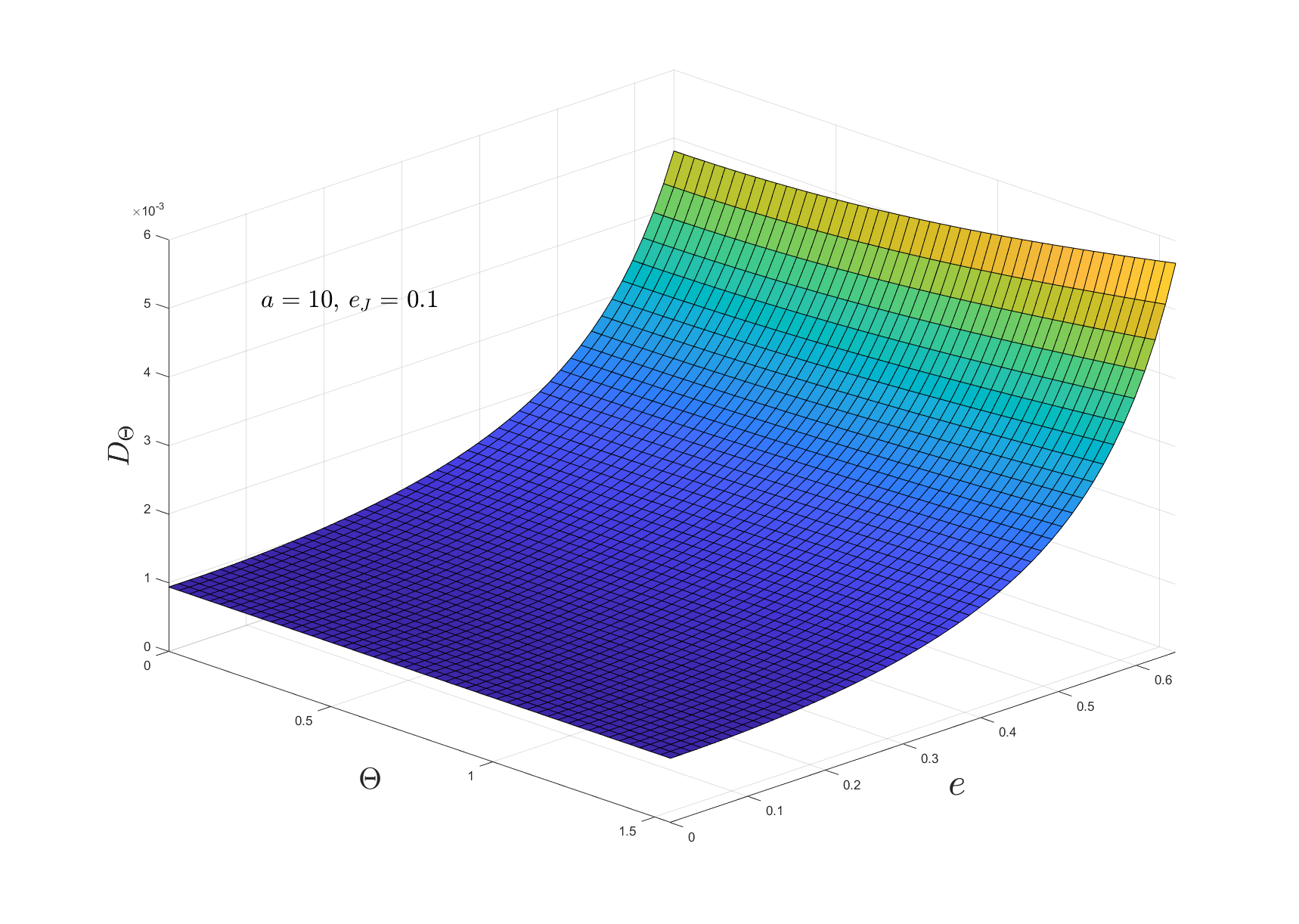}}
\caption{Value of $\bar{A}_{\Theta}$ and $D_{\Theta}$ for $a=10$ and $e_{J}=0.1$.}
\label{numeric4} 
\end{figure}

When $\Theta$ is not small, expansion of $\Theta$ can not be performed. Calculations of the coefficients of the quadratic form $\bar{W}_{\Theta}$ have been done numerically in Matlab. With the averaging procedure in (\ref{averageway}), detailing the expression of double-averaged values of coefficients in the quadratic form:
\begin{equation}
\begin{aligned}
\label{barABC E}
  & \bar{A}_{\Theta}=\frac{1}{{4{\pi}^{2} G}}
  \iint_{[0,2\pi]^2}
  \frac{y{{y}_{J}}\left( 1-e\cos  E  \right)\left( 1-{{e}_{J}}\cos {{E}_{J}}  \right)}{G {\left[ {\left( x-{{x}_{J}} \right)}^{2}+{{\left( y-{{y}_{J}} \right)}^{2}} \right]}^{{3}/{2}}}  \text{d}E\text{d}{E}_{J},\\
  & \bar{B}_{\Theta}=\frac{1}{{8{\pi }^{2} G}}
 \iint_{[0,2\pi]^2}
  \frac{\left( x{{y}_{J}}+y{{x}_{J}} \right)\left( 1-e\cos  E  \right)\left( 1-{{e}_{J}}\cos {{E}_{J}}  \right)}{G {\left[ {\left( x-{{x}_{J}} \right)}^{2}+{{\left( y-{{y}_{J}} \right)}^{2}} \right]}^{{3}/{2}}}  \text{d}E\text{d}{E}_{J},\\
   & \bar{C}_{\Theta}=\frac{1}{{4{\pi }^{2} G}}
  \iint_{[0,2\pi]^2}
  \frac{x{{x}_{J}}\left( 1-e\cos  E  \right)\left( 1-{{e}_{J}}\cos {{E}_{J}}  \right)}{G {\left[ {\left( x-{{x}_{J}} \right)}^{2}+{{\left( y-{{y}_{J}} \right)}^{2}} \right]}^{{3}/{2}}}  \text{d}E\text{d}{E}_{J}.
 \end{aligned}
\end{equation}
Taking $a_{J}=1$, for chosen values of $a$ and $e_{J}$, each pair of $\left(\Theta, e \right)$ belongs to one certain periodic orbit which is determined by
\begin{equation}
\label{bar_R}
\bar R =-\frac{1}{{4{\pi}^{2}}}\iint_{[0,2\pi]^2}{{\frac {\left( 1-e\cos  E  \right)\left( 1-{{e}_{J}}\cos {{E}_{J}}  \right) }{\sqrt{(x-x_{J})^{2}+(y-y_{J})^{2}}}}} \text{d}E\text{d}{E}_{J}.
\end{equation}
Considering values of variables satisfying conditions (\ref{condition}), and then plotting the figures of $(\Theta, e, \bar{A}_{\Theta})$ and  $(\Theta, e, D_{\Theta})$ numerically for some considerable $a$ and $e_{J}$. The numerical work takes $a=0.3$ and $e_{J}=0.2$ in Fig. \ref{numeric1}, $a=0.2$ and $e_{J}=0.4$ in Fig. \ref{numeric2}, $a=4$ and $e_{J}=0.4$ in Fig. \ref{numeric3}, $a=10$ and $e_{J}=0.1$ in Fig. \ref{numeric4}. All the numerical results demonstrate that the sequential principal minor $\bar{A}_{\Theta}>0$ and $D_{\Theta}>0$, thus the quadratic form $\bar{W}_{\Theta}$ is positive defined. Therefore, the numerical calculation gives stability of the periodic orbits of the asteroid in the linear approximation in the double-averaged spatial perturbed elliptic restricted 3-body problem for all values of parameters. 
%we choose some periodic orbits around the equilibria,  take $\Theta$ on some grid and $e$ is obtained from equations (\ref{for_e_*}), (\ref{for_e_*_1}).  we calculate $a$ from 0.1 to 10 and $e_{J}$ from 0.01 to 0.99 with some steps. Then, we choose a $\Theta$ from $0$ to $\pi$ and a $e$ from 0 to 1, 


\section{Conclusion}

A complete analysis of secular effects on the motion of a massless asteroid within the framework of the spatial perturbed, double-averaged elliptic restricted three-body problem has been conducted. The stability of the asteroid's orbits was investigated with linearization. Notably, periodic orbits originating from the planar problems are stable within the spatial perturbation across all parameter regions. Numerical simulations corroborate these findings for a variety of periodic orbits. The model's applicability to systems with highly eccentric planets renders it particularly valuable for exoplanet studies.

\section*{Acknowledgements}

The authors express their gratitude to Prof. Anatoly Neishtadt for suggestions on the topic and to Prof. Xijun Hu for discussions. Kaicheng Sheng thanks the National Natural Science Foundation of China (NSFC) for the support of this research (Grant:  12371192 \& 12271300). 

%\section*{Data availability}

%Data sharing does not apply to this article as no new data were created or analyzed in this study.

%\section*{Conflict of interest statement}

%The authors declare that there are no conflicts of interest, we do not have any possible conflicts of interest.

\bibliographystyle{plainnat}  
\bibliography{reference}

\end{document}